\documentclass[useAMS,usenatbib]{mn2e}

\usepackage{epsfig}
\newcommand{\kpc}{\mbox{\rm kpc}}   
\newcommand{\Gyr}{\mbox{\rm Gyr}}   
\newcommand{\msun}{\mbox{$M_\odot$}} 
\newcommand{\OH}{{\rm[O/H]}}
\newcommand{\FeH}{{\rm[Fe/H]}}
\newcommand{\OFe}{{\rm[O/Fe]}}
\newcommand{\MH}{\mbox{$M_{\rm H}$}}
\newcommand{\MO}{\mbox{$M_{\rm O}$}}
\newcommand{\MFe}{\mbox{$M_{\rm Fe}$}}

\title[Star Formation and Metallicity in Barred Spiral Galaxies]
{The Connection between
Star Formation and Metallicity Evolution in Barred Spiral Galaxies}
\author[Hugo Martel et al.]
{Hugo Martel,$^{1,2}$ Daisuke Kawata,$^3$ and
Sara L. Ellison$^4$\\
$^{1}$D\'epartement de physique, de g\'enie physique et d'optique,
Universit\'e Laval, Qu\'ebec, QC, G1V 0A6, Canada\\
$^{2}$Centre de Recherche en Astrophysique du Qu\'ebec,
C.P. 6128, Succ. Centre-Ville, Montr\'eal, QC, Canada\\
$^3$Mullard Space Science Laboratory, University College London,
Holmbury St. Mary, Dorking, Surrey, UK\\
$^4$Department of Physics and Astronomy, University of Victoria,
Victoria, BC, Canada}
\begin{document}

\date{Accepted XXX. Received XXX; in original form XXX}

\pagerange{\pageref{firstpage}--\pageref{lastpage}} \pubyear{XXX}

\maketitle

\label{firstpage}

\begin{abstract}
We have performed a series of chemodynamical simulations of barred
disc galaxies. Our goal is to determine the physical processes
responsible for the increase in the central gas-phase metallicity and
of the central star formation rate (SFR) observed in the {\sl Sloan Digital
Sky Survey} (SDSS). All simulations start with an axisymmetric
distribution of stars and gas, embedded into a spherical dark matter
halo. We define a $2\,\kpc$ diameter central aperture to approximate the
integrated spectroscopic fibre measurements from the SDSS. The
chemical evolution observed within this central region depends
critically upon the relative size of the bar and the aperture, which
evolves strongly with time.  At $t\sim0.5\,\Gyr$, a strong bar forms via a
disk instability, whose length is considerably longer than the $2\,\kpc$
aperture. The stars and gas lose angular momentum and follow elongated
orbits that cause an intense mixing of the gas between the central
region and its surroundings. During the next $1.5\,\Gyr$, the orbits of
the stars inside the bar do not evolve much, but the orbits of the gas
contract significantly until the entire gas bar is contained in the
$2\,\kpc$ aperture, resulting in a net flux of gas into the
central region. During this period, the metallicity in the central
region increases steadily, and this enrichment is dominated by
metal-rich gas that is flowing into the central region. The main result of
this work is therefore that the observed enrichment in the centres of
barred galaxies is not dominated by in-situ enrichment by stars formed
in the centre. Rather, star formation occurs along the full length of
the bar, much of which occurs initially outside the $2\,\kpc$ aperture.
About 50\% of the metals that end up in the central region originate
from this extended bar-long star formation, but flow into the central
region due to loss of angular momentum. The effect is less
significant for iron because the delay for the onset of Type-Ia SNe
leaves less time for mixing. Still, there is a significant increase
in \FeH\ before the central stars contribute to in-situ enrichment.
Eventually, as the orbits of the gas inside the bar contract, they
fall completely inside the central region, and only then the central
region can be regarded as a closed-box system. However, by that time,
most of the metal-enrichment in the central region has already taken
place. We conclude that there is no direct connection between central
SFR and central metallicity. The central metallicity does not
originate exclusively from central stars. Instead, the global SFR
(especially along the bar) and the large-scale flow of enriched gas
play a major role.
\end{abstract}


\section{Introduction}

\subsection{Evolution of Barred Galaxies}

Bars are a very common feature in spiral galaxies. Observational
determinations of the bar fraction
in nearby galaxies, and up to redshift $z\sim1$, systematically produce
values in the range 20\%
to more than 60\% \citep{eskridgeetal00,eeh04,jogeeetal04,na10,mm11,gtms11,
mastersetal11,leeetal12a}. The formation of these bars can
be triggered by mergers or tidal interactions with other galaxies
\citep{noguchi87}, or can
happen in isolated galaxies as a result of dynamical instability
\citep{sellwood81}. 
Analytical studies and numerical simulations have shown that the presence of
a bar can greatly affect the evolution of the host galaxy.
Analytical calculations \citep{lyndenbell79,athanassoula03}
showed that the bar produces a gravitational torque that transports angular
momentum outward
and matter inward. This was confirmed by the early simulations of 
\citet{rhva79}, \citet{sellwood81}, and \citet{var81}.
In the more recent simulations of \citet{pf91},
\citet{fb93,fb95} and \citet{fbk94},
a bar forms by dynamical instability. That bar channels gas into the center
of the galaxy, resulting in a flattening of the initial metallicity gradient.
Eventually, the gas flow causes a central starburst, and the resulting
metal enrichment steepens the metallicity gradients in the central regions.
Several authors studied the dynamics of the gas flowing along the bar
\citep{cg85,athanassoula92b,ce93,fux99,mtss02,rt04,bsw10}.
Their simulations showed that the gas
retains some angular momentum, preventing it from falling directly
into the center of the galaxy. Instead, the gas tends to form
elongated orbits inside the stellar bar. This result is in agreement with
analytical studies of the properties of orbits in axisymmetric potentials
\citep{athanassoula92a}. If gas does fall into the center of the galaxy,
it might accrete onto a central black hole and fuel a central AGN 
\citep{sfb89,sn93,hs94,combes03,jogee06}. Or, it might lead
to the formation of a central bulge \citep{kk04}.

All these analytical studies and numerical simulations provide a basic
scenario for the evolution of barred galaxies. Once the stellar bar forms, gas
starts flowing along the bar, into the central regions. 
This inflow of metal-poor gas tends to flatten the metallicity gradient.
The star formation
rate in the central regions increases, either steadily or in the form
of a starburst. This will lead eventually to an increase in central
metallicity. Finally, the gas that is not converted to stars might end
up fueling a central AGN.
We would therefore expect barred galaxies to have higher central SFR,
higher central metallicities, higher bulge fraction,
and more intense AGN activity than
unbarred galaxies, and possibly a flatter metallicity gradient,
depending at what stage of their evolution they are being observed. 
This basic scenario is indeed supported by many observations.
Some barred galaxies have shallower metallicity gradients than unbarred ones
\citep{pageletal79,mr94}, suggesting gas flow along the bar. 
Several observations reveal enhanced star formation
in the central regions of barred galaxies
\citep{heckman80,hawardenetal86,devereux87,arsenault89,
huangetal96,hfs97,mf97,hm99,emsellemetal01,knapenetal02,lsb04,jogeeetal05,
huntetal08,ba09}. 
Also, \citet{mastersetal12} found an anti-correlation between the
presence of a bar and atomic gas content, which could be interpreted
as a depletion of gas in barred galaxies, possibly resulting from
enhanced star formation. \citet{mastersetal11} found that over half of red,
bulge-dominated disc galaxies possess a bar, while blue galaxies show no 
evidence for a bar or a bulge.
The issue of AGN fueling is less clear. Several
observational studies find either a larger bar fraction 
in AGN-host galaxies than 
non-AGN ones \citep{arsenault89,knapenetal00,laineetal02} or
a larger AGN fraction in barred galaxies than unbarred ones
\citep{cg11},
other studies do not find any significant difference
\citep{mmp95,mr95,mr97,hfs97,lsb04,haoetal09,ba09,leeetal12b}.

\subsection{Recent Observations}

While many observations support the basic scenario of bar evolution described 
above,
several recent studies of barred galaxies using large samples reveal a
more complex picture. 
A study of 294 galaxies with strong bars from the {\sl Sloan Digital 
Sky Survey\/} (SDSS) by 
\citet{ellisonetal11} showed that barred galaxies of stellar masses
$M_\ast>10^{10}M_\odot$ have enhanced central SFR and
metallicities, which puts them as outliers on the fundamental metallicity
relation \citep{ellisonetal08,
mannuccietal10,laralopezetal10a,laralopezetal10b}.
Galaxies with stellar masses $M_\ast<10^{10}M_\odot$
also show an increase in central metallicity, but without
a corresponding increase in central SFR. One possible explanation is
that star formation 
in the center of low-mass barred galaxies, which is presumably responsible
for the observed higher metallicities, has now ceased, while stars are still
forming in the center of high-mass galaxies.

More recently, \citet{wangetal12} performed a detailed study of a 
sample of 3757 galaxies from SDSS, including 1555 barred galaxies. 
They found that the presence of a bar does not automatically 
imply higher SFR. Galaxies with weak bars 
(ellipticities $e_{\rm bar}<0.5$) do not show enhanced central star formation,
only galaxies with strong bars do.
These strong bars are found predominantly in galaxies with high stellar
masses, $M_\ast>1-3\times10^{10}M_\odot$ \citep{na10,ellisonetal11,wangetal12}.
This implies a relation between mass and SFR, which
could explain the results of \citet{ellisonetal11}. 
But interestingly, while only galaxies with strong bars in the 
\citet{wangetal12} sample
have enhanced central SFR, not all of them do.
Instead, some of these galaxies have a central
SFR that is lower than the mean by a factor of 10. This suggests that
bars (or some other process associated with their presence)
might, under some circumstances, quench star formation in the central
regions. 

\citet{ooy12} studied the effect of bars on central star formation 
and AGN 
activity, using a sample of 6658 late-type galaxies from SDSS, including
2442 galaxies with bars. They find that the presence of a bar significantly 
increases the central SFR in red galaxies, but not in blue ones, 
where the central SFR is large whether or not a bar is present.
This could result from the fact that redder galaxies tend to have
longer bars, which according to simulations leads to larger effects. 
Or, as the authors suggest, central star formation is naturally 
reduced in redder galaxies because less gas is available, making the effect of
infalling gas more important than in bluer galaxies. While bar effects
on the central SFR are important for red galaxies but not blue ones, {\it the
opposite is true for AGN activity\/}. Bar effects on AGNs are mostly found
in blue galaxies, even though most AGNs are hosted by red galaxies.
The authors suggest that the presence of a large central mass concentration
can weaken the bar, reducing its effects. Another possibility
is that red galaxies
hosting AGNs are in the post-starburst phase. Then most of the gas has already 
been consumed by star formation, leaving no gas to feed the AGN
\citep{shethetal05}. 

These recent studies reveal that the evolution of barred galaxies
is probably more complex than the basic scenario suggests. Galaxies
of different masses, or with bars of different strength, can have drastically
different star formation histories. As a result, there is no simple correlation
between the presence of bars, enhanced central SFR, and enhanced 
central metallicity. To fully
understand the evolution of these galaxies, we need to understand the 
interplay between the gravitational dynamics responsible for the formation of
the bar and the large-scale gas flows in the galaxy, the hydrodynamics 
responsible for the formation of stars and the chemical enrichment of the 
ISM, the feedback processes resulting from stellar evolution and AGN 
activity, and their possible role in dispersing chemical elements in the
galaxy, regulating star formation, and possibly quenching it.

\subsection{Objectives}

To address this complex problem, we first focus, in this paper,
on the connection between enhanced central star formation and enhanced
central metallicity \citep{ellisonetal11}. 
This is a first step toward understanding the evolution
of barred galaxies and building a scenario which is consistent with all
recent observations.
Previous hydrodynamical simulations of barred galaxies showed that 
gas flow along the bar usually results in increased star formation and
increased metallicity in the central regions \citep{pf91,fb93,fb95,fbk94},
but that does not imply that
one is the cause, or partial cause, of the other. Metals that end up in
the central regions might have been produced elsewhere in the galaxy, while
metals produced in the central regions might eventually leave that region.
These specific questions were not addressed in previous simulations.

We have performed a series of chemodynamical simulations of isolated
barred galaxies.
We focus on the evolution of the metal abundance in the central region. 
We define this central region as
the central $2\,\kpc$, because it roughly matches the size of the
fibre region in the SDSS. From now on, we will use the term
``central,'' to refers to the $2\,\kpc$ aperture, except in section 3.6 where 
we consider different aperture sizes.
We identify all physical processes responsible for modifying the
abundance of individual chemical elements in the central region, and 
we determine
the relative importance of these various processes. Our objective is to
determine if enhanced central star formation is the primary cause for the
enhanced central metallicity, or if the latter has a different origin.
This is an essential first step in assessing the validity or
shortcomings of the basic scenario. 

The remainder of this paper is organized as follows: In \S2, we describe 
the numerical algorithm used for this study. Results are presented in \S3,
where we first discuss the global evolution of the galaxy, and then
focus on the central region. The implications of the results are discussed
in \S4. Summary and conclusions are presented in \S5.

\section{THE SIMULATIONS}

\subsection{The Numerical Algorithm}

We used the chemodynamical numerical algorithm
GCD+ \citep{kg03,rk12,kawataetal13}. 
GCD+ is a three-dimensional tree N-body/SPH code that 
incorporates self-gravity, hydrodynamics, radiative cooling, star formation, 
supernova feedback, metal enrichment, and metal diffusion. 
Gas is converted into stars in regions where the number density
exceeds a certain threshold $n_{\rm th}$ and where the
velocity field is convergent, following the Schmidt law described in
\citet{kg03}:
\begin{equation}
{d\rho_*\over dt}=-{d\rho_g\over dt}={C_*\rho_g\over t_g}
\end{equation}

\noindent
where $\rho_g$ and $\rho_*$ are the gas and stellar mass density, respectively,
$C_*$ is a dimensionless SFR efficiency, and $t_g$ is the dynamical time.

We assume that the the stellar masses are distributed according to the 
\citet{salpeter55} initial mass function (IMF). Chemical enrichment by both 
Type~II \citep{ww95} and Type~Ia supernovae \citep{iwamotoetal99} and mass
loss from intermediate-mass stars \citep{vdhg97} are taken into account.
We adopt the Type Ia supernovae rate suggested by \citet{ktn00}. 
We assume that each supernova produces an amount of thermal energy 
$E_{\rm SN}$, and
that stellar winds from massive stars ($M>30\msun$) also produce 
thermal energy, at a rate $\dot E_{\rm SW}$. The canonical energy of a
supernova is $10^{51}\rm erg$. However, 
according to the high-resolution one-dimensional simulations
of \citet{thorntonetal98}, 90\% of the initial supernova 
energy is lost in radiation in its early expansion phase, and does not
contribute to feedback.
Accordingly, we set $E_{\rm SN}=10^{50}\rm erg$ per 
supernova for our fiducial model.
Notice that an initial supernova energy has not
been established quantitatively yet. For this reason, we 
also explore the effect of a higher supernova energy, 
$E_{\rm SN}=10^{51}\rm erg$, in one of our 
simulations. We set $\dot E_{\rm SW}=10^{36}\rm erg\,s^{-1}$
for our fiducial model, but also consider
the effect of a stronger stellar wind with 
$\dot E_{\rm SW}=10^{37}\rm erg\,s^{-1}$, to cover the range suggested by several
authors \citep{weaveretal77,shull80,gibson94,om94}. 
The metal diffusion is modeled with the method described in \citet{ggbk09}.
We use a star formation density threshold, 
$n_{\rm th}=0.1\,{\rm cm}^{-3}$ and a star formation efficiency, $C_*=0.02$.

\subsection{Initial Conditions}

The technique used for generating initial conditions is described
in detail in \citet{gkc12}. We use particles to represent the stellar and
gaseous components of the galaxy.
The stellar and gas discs are set up using the method described in 
\citet{smh05}. The stellar disc follows an exponential surface density profile:
\begin{equation}
\rho^{\phantom1}_{\rm d,stars}=
{M_{\rm d,stars}\over4\pi z^{\phantom1}_{\rm d,stars}R_{\rm d,stars}^2}
\,e^{-R/R_{\rm d,stars}}\,{\rm sech}^2{z\over z^{\phantom1}_{\rm d,stars}}\,,
\end{equation} 

\begin{table}
 \centering
 \begin{minipage}{80mm}
  \caption{Numerical Parameters of the Initial Conditions}
  \begin{tabular}{@{}lccc@{}}
  \hline
     Disc & $M_{\rm d}$ [$\msun$] & $R_{\rm d}$ [$\kpc$] & $z_{\rm d}$ [$\kpc$] \\
  \hline
     stars & $4\times10^{10}$ & 2.50 & 0.35 \\
     gas   & $1\times10^{10}$ & 4.00 & $\cdots$ \\
   \hline
\end{tabular}
\label{discs}
\end{minipage}
\end{table}

\noindent
where $R$, $z$, $M_{\rm d}$, $R_{\rm d}$ and $z^{\phantom1}_{\rm d}$,
are the radial coordinate (distance from the $z$-axis), vertical coordinate,
disc mass, scale length, and scale height, respectively. 
The gaseous disc has a same radial dependence as the stellar disc,
but with a different scale length $R_{\rm d,gas}$,
and its initial vertical distribution is not given by a ${\rm sech}^2z$ profile.
Instead, it is adjusted by imposing that the
gas is initially in hydrostatic equilibrium.
The values of the parameters are given in Table~\ref{discs}.
These parameters were chosen in order
to produce a galaxy similar to the Milky Way
\citep{om01,robinetal03,mcmillan11}.
In this first study, we did not include gas-infall from the surrounding
intergalactic medium, in order to focus on the effect of gas accretion
in the disc.

We impose initial radial abundance profiles to both stellar and gaseous
components. The iron abundance profiles for the stars and gas are given by
\begin{equation}
\FeH(R)=0.2-0.05\left({R\over1\,\kpc}\right)\,.
\label{FeH}
\end{equation}

\noindent
The $\alpha$-element abundances are given by
\begin{equation}
[\alpha/{\rm Fe}](R)=\cases{-0.16\FeH\,,& stars\,;\cr
0\,,& gas\,.\cr}
\label{aFe}
\end{equation}

\noindent At each radius, we add to the values given by 
equations~(\ref{FeH}) and (\ref{aFe}) a gaussian scatter with a
dispersion of $0.02\,{\rm dex}$. For the star particles, we use
the value of \FeH\ to randomly assing them an age, following the
age-metallicity relation $\FeH=-0.04\times\hbox{age (Gyr)}$ with a
0.02~dex scatter, and assuming a constant SFR in the $0-10\,\rm Gyr$
age range.

We do not use particles to represent the 
dark matter halo. Instead, we impose a fixed, non-evolving spherical
gravitational potential, in order to focus on the baryon dynamics
and save computational time
The dark matter halo density follows a truncated NFW profile,
\begin{equation}
\rho_{\rm dm}={\rho_{\rm c}\over (r/r_s)(1+r/r_s)^2}e^{-(r/r_{200})^2}
\label{eq1}
\end{equation} 

\noindent \citep{nfw97,ras09}
where $r$ is the distance from the center.
The scale length $r_{\rm s}$ is given by $r_{200}/c$, where $c$ is the
concentration parameter, and $r_{200}$ is the virial radius, 
inside which the mean
density of the dark matter is equal to 200 times the
the critical $\rho_{\rm crit}$. The characteristic density
$\rho_{\rm c}$ is adjusted such that the mass $M_{200}$ of the dark matter
inside the virial radius is equal to $2\times10^{12}\msun$.
The truncation factor $\exp(-r^2/r_{200}^2)$ is introduced in our initial
condition generator for a live halo simulation. Although we use a static
dark matter halo in this paper, we still use the profile given by 
equation~(\ref{eq1}). Note that adding this truncation factor to the
standard NFW profile leads to
very little change in the central region, which we focus on in this paper.

\subsection{Runs and Parameters}

\begin{table}
 \centering
 \begin{minipage}{80mm}
  \caption{Numerical Parameters of the Simulations}
  \begin{tabular}{@{}lccccc@{}}
  \hline
    Run   & $N_{\rm stars}$ & $N_{\rm gas}$ & $c$ & $E_{\rm SN}$ [${\rm erg}$]
          & $\dot E_{\rm SW}$ [${\rm erg\,s^{-1}}$] \\
  \hline
    F     & 120,000 & 30,000 & 10 & $10^{50}$ & $10^{36}$ \\
    HR    & 360,000 & 90,000 & 10 & $10^{50}$ & $10^{36}$ \\
    NoBar & 120,000 & 30,000 & 20 & $10^{50}$ & $10^{36}$ \\
    Feed  & 120,000 & 30,000 & 10 & $10^{51}$ & $10^{37}$ \\
  \hline
\end{tabular}
\label{runs}
\end{minipage}
\end{table}

We performed a total of 4 simulations. The parameters of the
simulations are listed in Table~\ref{runs}.
We first performed a {\it fiducial run\/}, or Run~F, with a medium resolution
(120,000 star particles and 30,000 gas particles), concentration parameter 
$c=10$, supernova energy input $E_{\rm SN}=10^{50}\,{\rm erg}$ per supernova,
and stellar wind energy input $\dot E_{\rm SW}=10^{36}\,{\rm erg\,s^{-1}}$.
We then performed a
series of simulations which all differ from the fiducial run in one particular
aspect: Run HR is a high-resolution run, with three times more particles.
In Run NoBar, we increased the concentration parameter of the dark matter
halo, to prevent the formation of a bar. In Run Feed, 
we increased amount of stellar wind and SNe feedback by a factor of 10.
All simulations end at $t=2\,\Gyr$.

\section{RESULTS}

\subsection{Global Properties}

\subsubsection{The Formation and Evolution of the Bar and Spiral Pattern}

\begin{figure*}
\begin{center}
\includegraphics[width=6.5in]{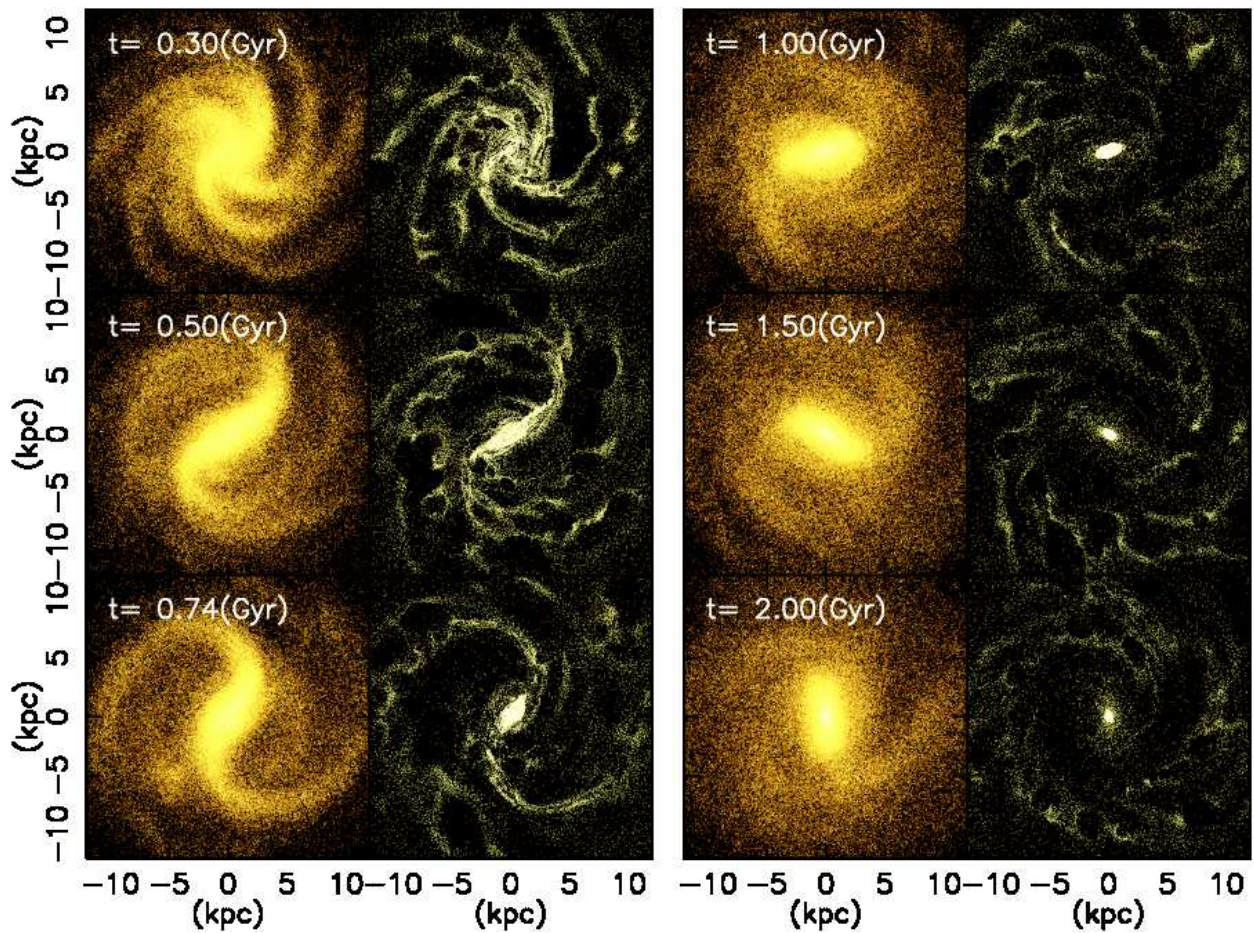}
\caption{Distribution of star particles (left panel in each pair)
and gas particles (right panel in each pair) for Run F
at six different times, as indicated. Each panel is $25\,\kpc\times25\,\kpc$
in size.}
\label{xy}
\end{center}
\end{figure*}

Figure~\ref{xy} shows the distribution of stars and gas particles at six
different times, for Run F. At $t=0.30\,\Gyr$, we
see a spiral pattern with multiple spiral arms, and a central bar is 
starting to form. At $t=0.50\,\Gyr$, the central bar is fully developed,
and the spiral pattern is reduced to two main spiral arms. The stellar and 
gaseous components of the bar have a comparable length at that time.
By $t=1\,\Gyr$,
the bar is still present, but the spiral pattern starts to disappear
due to the dynamical heating by the bar and spiral arms themselves
\citep{cs85}.
At $t=1.5\,\Gyr$, the spiral pattern is hardly visible in the stellar
component, while in the gas component we see several weak spiral arms.
At the end of the simulation, the spiral pattern
is gone but there is still a strong bar in the stellar component.
Between $t=0.50\,\Gyr$ and the end of the simulation, the length and shape
of the stellar bar remain essentially unchanged, while the gaseous component
of the bar steadily contracts. At $t=2.00\,\Gyr$, the stellar and gaseous
components of the bar are $7\,\kpc$ and $1\,\kpc$ long, respectively.

\begin{figure*}
\begin{center}
\includegraphics[width=5in]{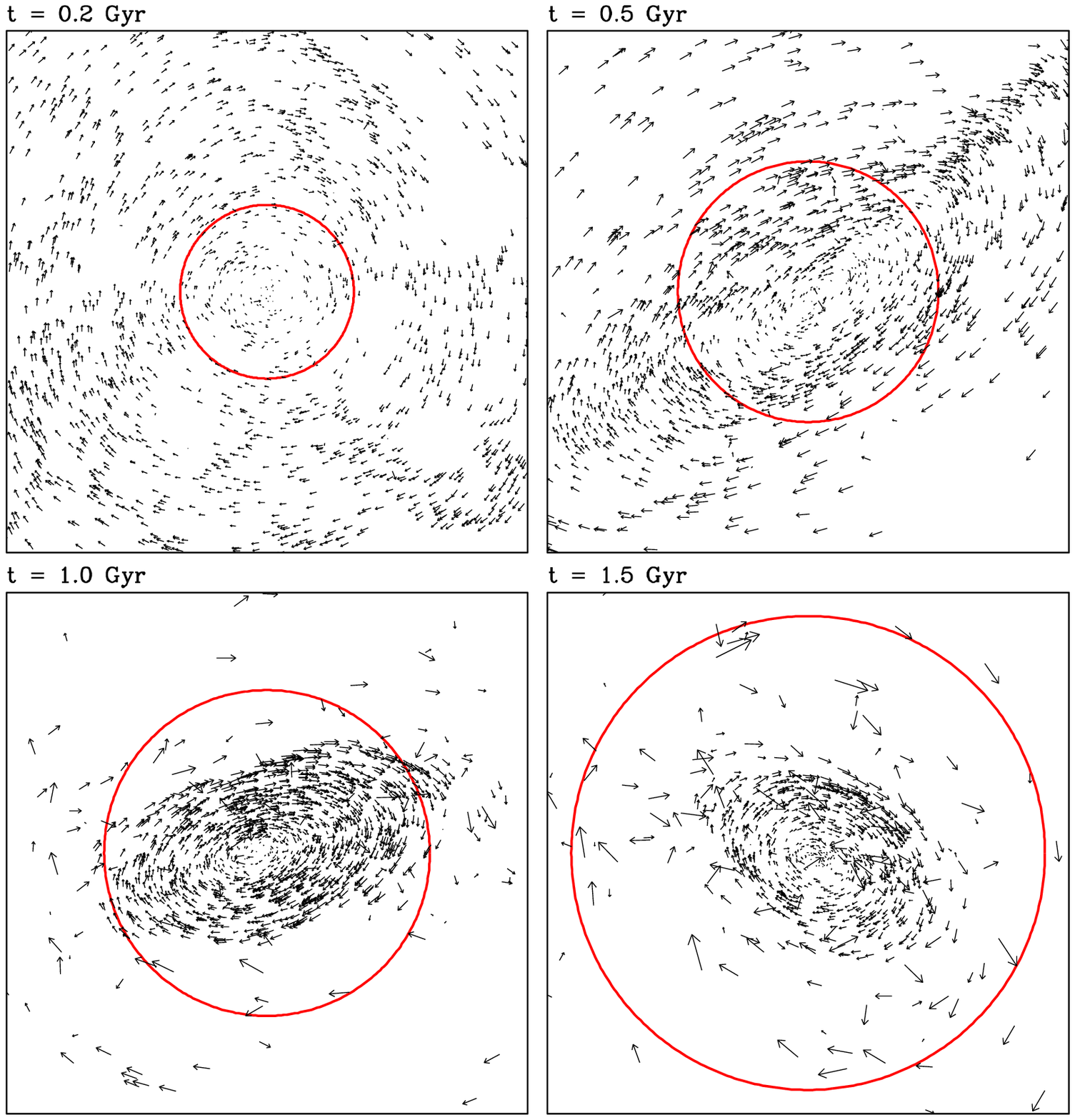}
\caption{Gas velocity in the central region for Run F
at four different times, 
as indicated. The red circle shows the central region, and has a diameter of 
$2\,\kpc$. For clarity, only 1/4 of the particles are plotted.
}
\label{xy_vel}
\end{center}
\end{figure*}

We plot in Figure~\ref{xy_vel} the velocity field of the gas in the
central region. On each panel, the red circle has a diameter of
$2\,\kpc$, and shows the central region. At $t=0.5\,\Gyr$, the gas inside the
bar moves along an elongated, elliptical orbit. This is consistent with 
the simulations of \citet{rt04}. This orbit intersects the edge of the 
$2\,\kpc$ aperture, 
hence the gas is flowing across the boundary, both moving in and moving 
out of the simulated fibre.  This elongated vortex
contracts with time (see Fig.~\ref{xy}). At $t=1\,\Gyr$, the gas bar is barely 
contained within the $2\,\kpc$ aperture, and at
$t=1.5\,\Gyr$, it is entirely contained inside the central region.

\subsubsection{Star Formation and Chemical Evolution of the Galaxy}

\begin{figure*}
\begin{center}
\includegraphics[width=5in]{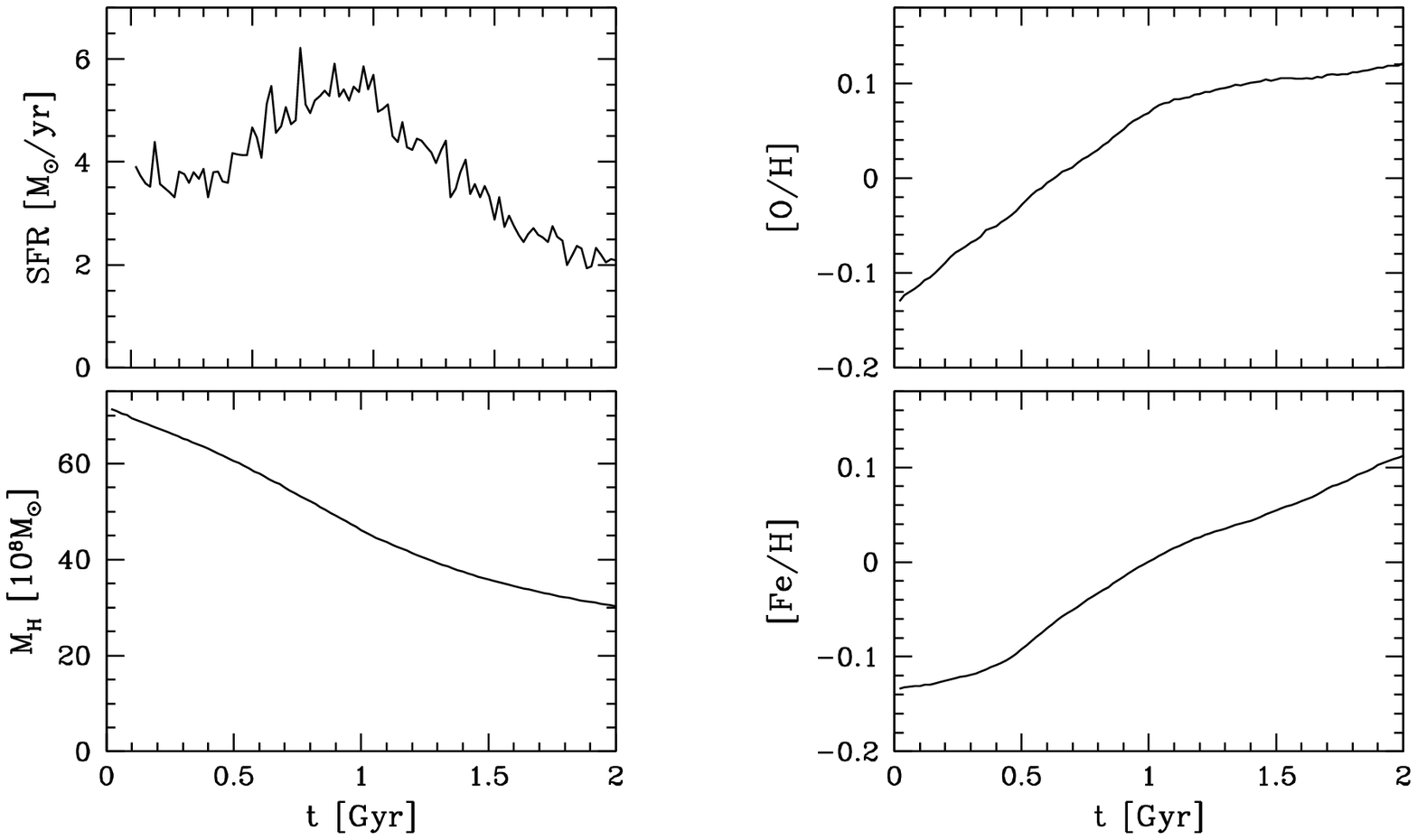}
\caption{Time-evolution of global quantities, for Run F.
Top left panel: Star formation rate; bottom left panel: total hydrogen mass
in the gas phase;
right panels: \OH\ and \FeH\ in the gas phase.}
\label{global}
\end{center}
\end{figure*}

Figure~\ref{global} shows the time-evolution of global properties of
the galaxy. The top left panel shows the SFR. The total galactic SFR
is initially constant at around $4\msun\,{\rm yr}^{-1}$.
Then, at $t=0.5\,\Gyr$, the bar forms and starts chanelling gas
in the central region, resulting in additional star formation.
The SFR keeps increasing, reaching
$6\msun\,{\rm yr}^{-1}$ at
$t=1\,\Gyr$, and then starts decreasing.
The decrease in SFR after $t=1\,\Gyr$ is not due
to bar disruption, since the
bar remains strong until the end of the simulation.
It is simply due to gas exhaustion: star formation reduces the amount of
gas available, and this gas is not replenished since we do not include
gas infall from the intergalactic medium in our simulations.
The bottom left panel shows the total hydrogen mass $\MH$ in the gas 
phase\footnote{Unless specified otherwise, elemental masses, abundances,
and abundances ratios always refer to the elements present
in the gas phase, not the
stellar phase. Also, we do not distinguish between molecular, atomic, and 
ion species. Hence, \MH\ refers to the total mass in the form of
$\rm H\,I$, $\rm H\,II$, and $\rm H_2$.}. 
It decreases almost linearly, with a slight change
of slope at $t=1.1\,\Gyr$. By the end of the simulations, it is down by 50\%,
and correspondingly, the SFR is also down by about 50\%

\begin{figure}
\begin{center}
\includegraphics[width=3.4in]{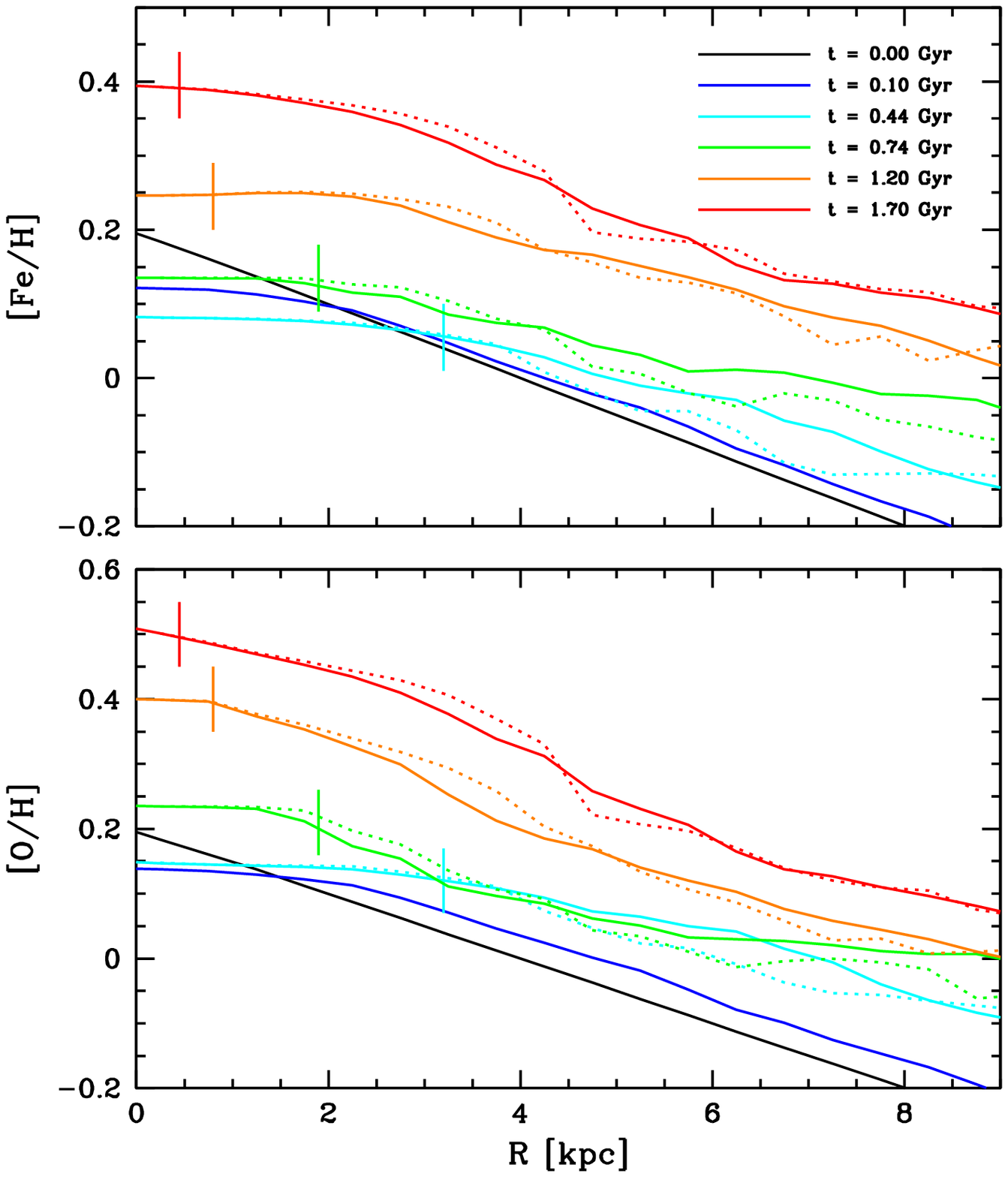}
\caption{Radial profiles of \OH\ and \FeH\ in the gas phase
at various epochs, for Run F.
Top panel: \FeH; bottom panel: \OH.
Colors represent different epochs, as indicated.
Solid lines: azimuthally-averaged profiles; dotted lines: profiles calculated
in a cylinder of radius $1\,\kpc$ centered on the bar.
Vertical lines indicate the half-length of the gaseous component of the bar,
determined by visual inspection.}
\label{profiles}
\end{center}
\end{figure}

The right panels show the evolution of 
\OH\ and \FeH. The evolution of \OH\ is closely related to
the evolution of the SFR.
The oxygen abundance rapidly increases during the period of 
high star formation, with \OH\ increasing by $0.21\,{\rm dex}$ during the
first $t\sim1.1\,\Gyr$. Then, as the SFR starts to decrease, there
is a significant change of slope, and \OH\ only increases by $0.04\,{\rm dex}$
during the last $0.9\,\Gyr$ of the simulation. The iron abundance increases
more steadily, and at the end of the simulation \FeH\ is increasing faster
than \OH. This is caused by the different lifetimes of Type II and Type Ia SNe
progenitors. Type II SNe have short progenitor lifetimes, between
5 and 10 million years. Since these SNe are the main producer of oxygen,
we expect to
see a tight correlation between the histories
of star formation and oxygen enrichment.
Iron is produced mostly by Type~Ia SNe, which have long progenitor lifetimes,
typically between 1 and $13\,\Gyr$.
As a result, a significant fraction of iron is produced long 
after the SFR has reached its peak. These results are consistent with the
early results of \citet{kg03} (see their Fig.~2).

Figure~\ref{profiles}
shows the evolution of the radial profiles of \OH\ and \FeH, 
where $R$ is the distance measured from the rotation
axis of the galaxy. We calculated these profiles first by averaging over
all angles, and then by only considering the gas located in a cylinder
of radius $1\,\kpc$
centered on the bar (solid and dotted lines, respectively).
The black lines show the linear gradient present in
the initial conditions. Once the bar forms, the flow of gas along the bar
tends to flatten the abundance gradient. In particular, \OH\ and \FeH\
in the central region initially go down, as metal-poor gas is channeled into
the central region by the bar. The same phenomenon was seen in the
simulations of \citet{fb93}, \citet{fbk94}, and \citet{fb95}.
This is partly compensated by metals produced by Type~II SNe. Since these
SNe produce more oxygen than iron, we see a larger drop in the central \FeH.
The SFR increases with time, resulting in more SNe, and 
eventually the oxygen and iron abundances increase with time at all radii.
The profiles tend to be flatter in the center when we consider only the
gas located along the direction of the bar. The length of the plateaux seen
at small radii in Figure~\ref{profiles} roughly correspond to the
extent of the gaseous component of the bar, which is indicate by the
vertical lines (see also Fig.~\ref{sfr_map2} below. Note that at 
$t=0.1\,\Gyr$, the bar has not formed yet).
Gas flow and strong mixing of gas inside the bar tends to erase any
significant abundance gradient. The plateaux get shorter with time simply 
because the gaseous component of the bar contracts.

\subsection{Inside the Central Region}

We define the {\it central region} as a cylinder of radius $1\,\kpc$, parallel
to the rotation axis of the galaxy, but centered on the center of mass of
the stellar component. We chose this particular size because it corresponds 
to the typical fibre values sampled by the SDSS.

\subsubsection{Star Formation and Metal Abundances}

Figure~\ref{fiber} shows the evolution of the SFR, hydrogen, oxygen, and
iron mass, and \OH, \FeH, and \OFe\ inside 
the central region. The dotted lines show the global evolution,
copied from Figure~\ref{global}. The chemical evolution of the central
region differs significantly from the global evolution of the whole galaxy.
Initially, the amount of hydrogen in the central region is small, and 
correspondingly the SFR is also small. As the bar starts channeling gas
toward the center, both $\MH$ and the SFR increase, especially after 
$t=0.5\,\Gyr$, when the bar is the strongest. 
At $t\sim1\,\Gyr$, the gaseous component of the bar has contracted
sufficiently to be entirely contained in the central region.
The inflow of hydrogen stops, and both $\MH$ and the SFR steadily 
decrease afterward. After $t\sim1.2\,\Gyr$,
star formation takes place almost exclusively in the central region, even
though only 1/4 of the available gas in the galaxy is located in
the central region at that time. The remainder of the gas is mostly located 
at large radii, inside the spiral arms, where the gas density is significantly
lower than in the central region.

In the middle left and bottom left panels
of Figure~\ref{fiber}, we also plotted the
oxygen mass \MO\ and iron mass \MFe\ inside the central region, after adjusting
the scales to match the evolution of \MH\ at early time. 
Before $t\sim0.6\,\Gyr$,
the masses of hydrogen and oxygen increase roughly at the same
{\it specific rate\/} $M^{-1}dM/dt$, and as a result the \OH\ 
remain fairly constant (top right panel). The same is true for iron,
up to $t\sim0.8\,\Gyr$ (see middle right panel). Afterward,
\MO\ and \MFe\ increase faster than \MH, causing the increase in 
\OH\ and \FeH. \MH\ reaches a 
peak at $t=0.85\,\Gyr$ while \MO\ and \MFe\ peak slightly later. After
reaching the peak, all masses decrease, but 
\MH\ decreases {\it faster\/} than \MO\ and \MFe, and consequently 
\OH\ and \FeH\ keep increasing.

The plots of \OH\ and \FeH\ show a similar behavior: there
is an initial drop in \OH\ and \FeH\ caused by the inflow of low-metallicity
gas into the center. This is partly compensated by enrichment from
Type~II SNe, which produce some iron and a lot more oxygen. 
At $t=1\,\Gyr$,
the SFR reaches its peak and starts decreasing. This reduces
the production of oxygen by Type II SNe, and causes a change of slope in \OH.
In the bottom right panel of Figure~\ref{fiber}, we plot
\OFe, which gives the relative importance of Type~II and Type~Ia
SNe. There is a clear change of regime at $t~\sim1.4\,\Gyr$, when
enrichment by Type~Ia SNe takes over.
Note that chemical enrichment is not entirely caused by stars formed during
the simulation. The stars present in the initial conditions contribute as well.
Therefore, enrichment by both Type~II and Type~Ia Sne starts at $t=0$,
following
the age distribution of the existing stars. Note also that Type~Ia SNe 
have a delay of about 0.4~Gyr from the birth of their
progenitors (see Fig.~1 of \citealt{ktn00}). 
As a result, when the SFR increases, the rate of Type~Ia SNe also increases
but with a 0.4~Gyr delay relative to the rate of Type~II SNe. 
After the SFR reaches its peak and
starts decreasing, the iron production, mainly due to Type~Ia SNe, can
exceed the oxygen production with a delay (depending on how quickly the SFR
decreases). Enrichment by Type~Ia SNe is more extended in time,
and as a result \FeH\ keeps increasing and \OFe\ keeps decreasing, at least
over the timescale of the simulation. However, note that we see an
initial increase of \OFe, because we set the initial \OFe\ to 0. 
If the initial value of \OFe\ was significantly higher, which would likely be
the case in high-redshift galaxies, we would not expect to see an initial 
upturn of \OFe\ due to the increase of SFR.

\subsubsection{Metallicity Evolution}

\begin{figure*}
\begin{center}
\includegraphics[width=5in]{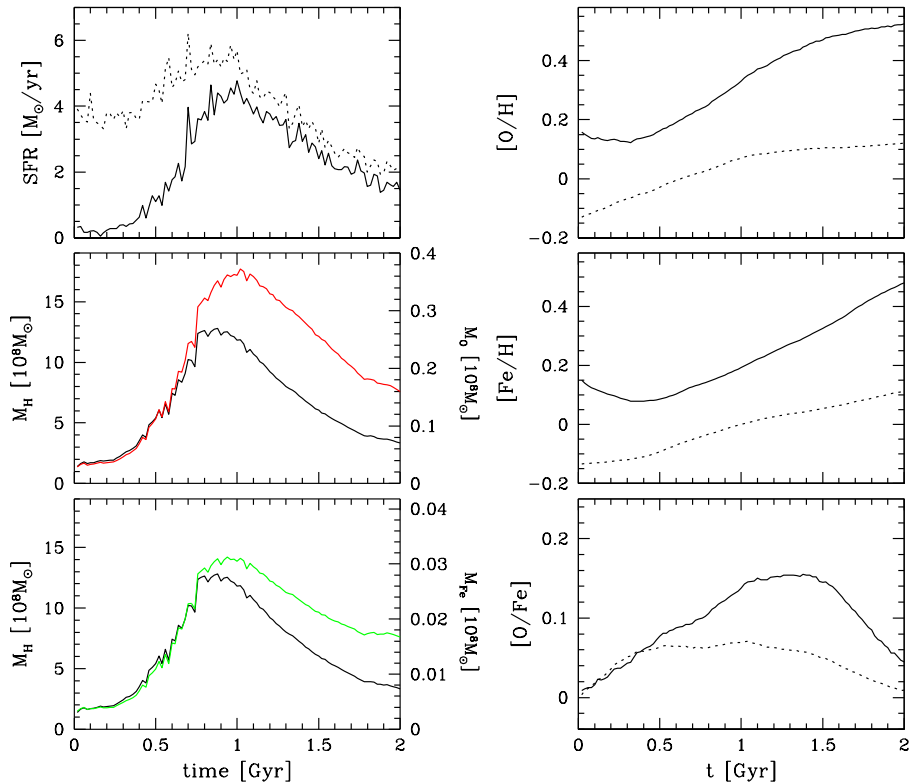}
\caption{Time-evolution of quantities inside the central region, for Run F.
Top left panel: Star formation rate (solid line);
middle and bottom left panels: hydrogen mass (black lines and left axes),
oxygen mass (red line and right axis in middle panel), and iron mass
(green line and right axis in bottom panel);
right panels: \OH, \FeH, and \OFe\ in the gas phase, 
as indicated (solid lines).
The dotted lines on four of the panels indicate the corresponding
global quantities (see Fig. \ref{global}).}
\label{fiber}
\end{center}
\end{figure*}

\begin{figure}
\begin{center}
\includegraphics[width=3.4in]{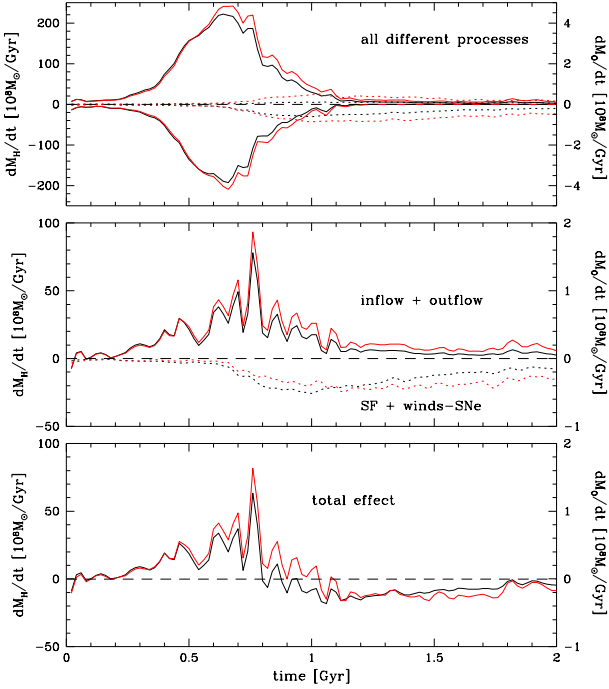}
\caption{Rate of change of hydrogen mass (black lines and left axes)
and oxygen mass (red lines and right axes) inside the central region,
versus time, for Run F. 
Top panel: solid lines show gas flowing into the central region 
(top lines)
and out of the central region (bottom lines). Dotted lines show gas added by
stellar winds and SNe outflows (top lines) and gas removed by star formation 
(bottom lines). Middle panel: sum of the quantities plotted in the
top panel. Solid lines show the net flux across the boundary of the
central region;
dotted lines show the net effect of matter interchange between the gas 
and stellar components. Bottom panel: sum of the quantities
plotted in the
middle panel, showing the net rate of change of mass, combining
all processes. Dashed lines indicate a constant mass.}
\label{fiber_OH2}
\end{center}
\end{figure}

\begin{figure}
\begin{center}
\includegraphics[width=3.4in]{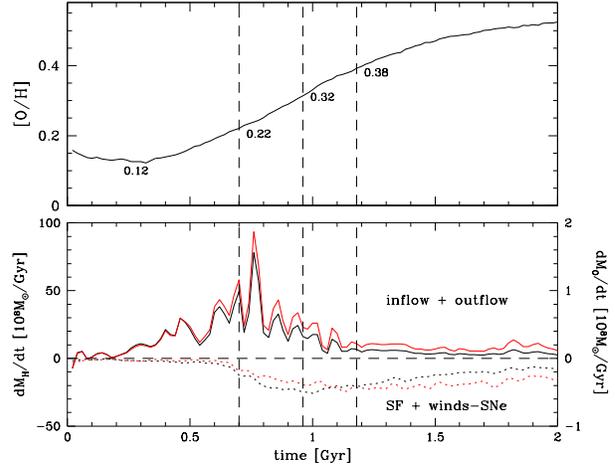}
\caption{Top panel: \OH\ in the gas phase
in the central region, versus time, for Run F.
Bottom panel:
rate of change of hydrogen mass (black lines and left axis)
and oxygen mass (red lines and right axis) inside the central region,
versus time. These panels are copied from Figures~\ref{fiber}
and~\ref{fiber_OH2}, respectively. Vertical dashed lines identify
various characteristic epochs. Numbers in top panel indicate
values of \OH.}
\label{fiber_2p}
\end{center}
\end{figure}

\begin{figure}
\begin{center}
\includegraphics[width=3.4in]{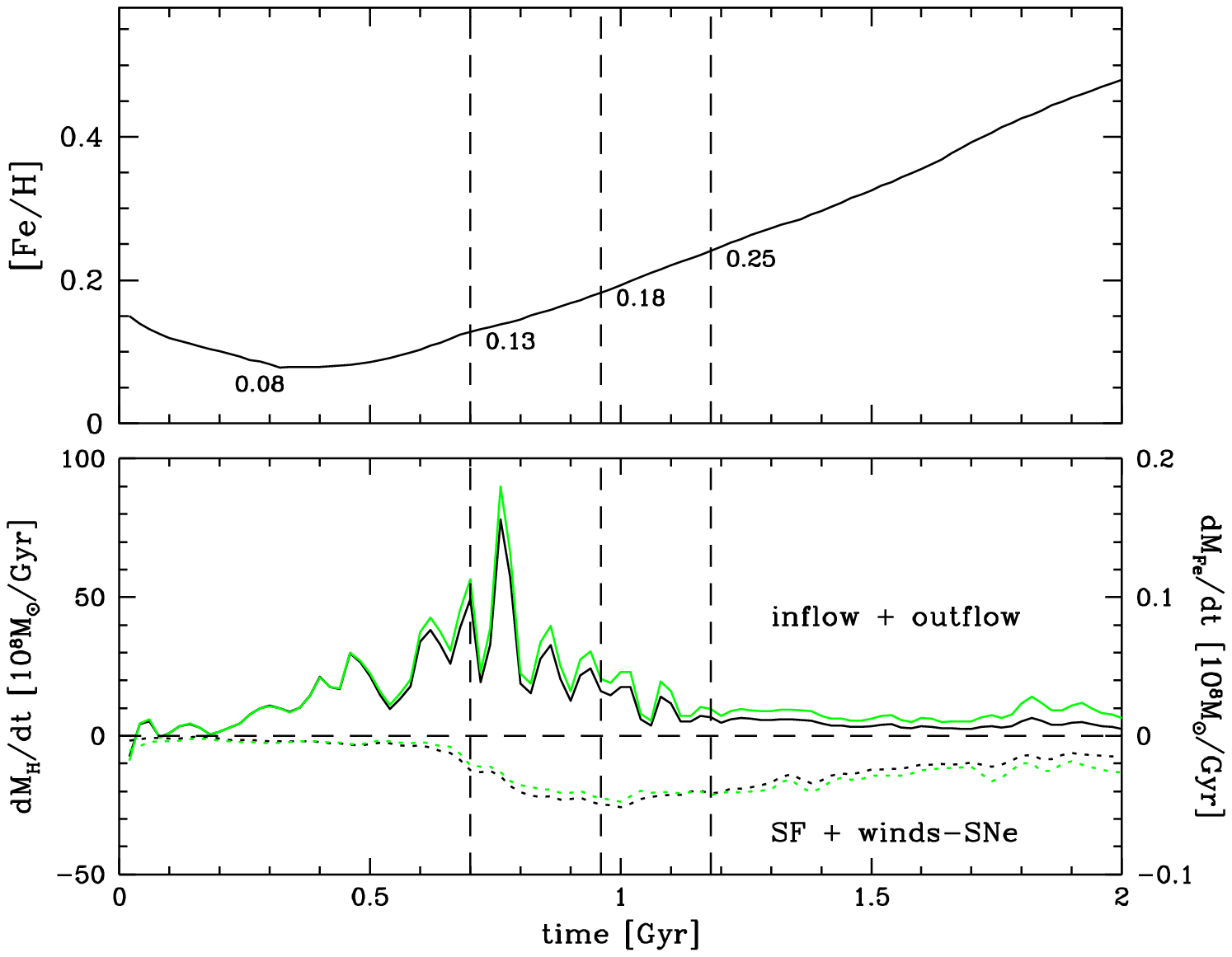}
\caption{Top panel: \FeH\ in the gas phase in the central region, versus time,
for Run F.
Bottom panel:
rate of change of hydrogen mass (black lines and left axis)
and iron mass (green lines and right axis) inside the central region,
versus time. Vertical dashed lines identify
various characteristic epochs. Numbers in top panel indicate values of \FeH.}
\label{fiber_Fe_2p}
\end{center}
\end{figure}

\begin{figure}
\begin{center}
\includegraphics[width=3.4in]{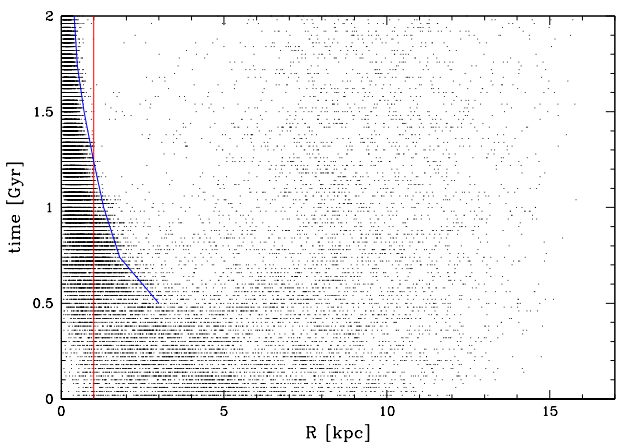}
\caption{Star formation history for Run F. Each dot
corresponds to a star formation event (a gas particle turning into a
star particle). The figure shows the epoch of the event versus radius.
The red line shows the boundary of the central region. The blue line
shows the half-length of the gaseous component of the bar, determined by
visual inspection. The dots located on the right side of that curve 
correspond to stars formed in the disc, mostly inside the spiral arms.}
\label{sfr_map2}
\end{center}
\end{figure}

\begin{figure}
\begin{center}
\includegraphics[width=3.4in]{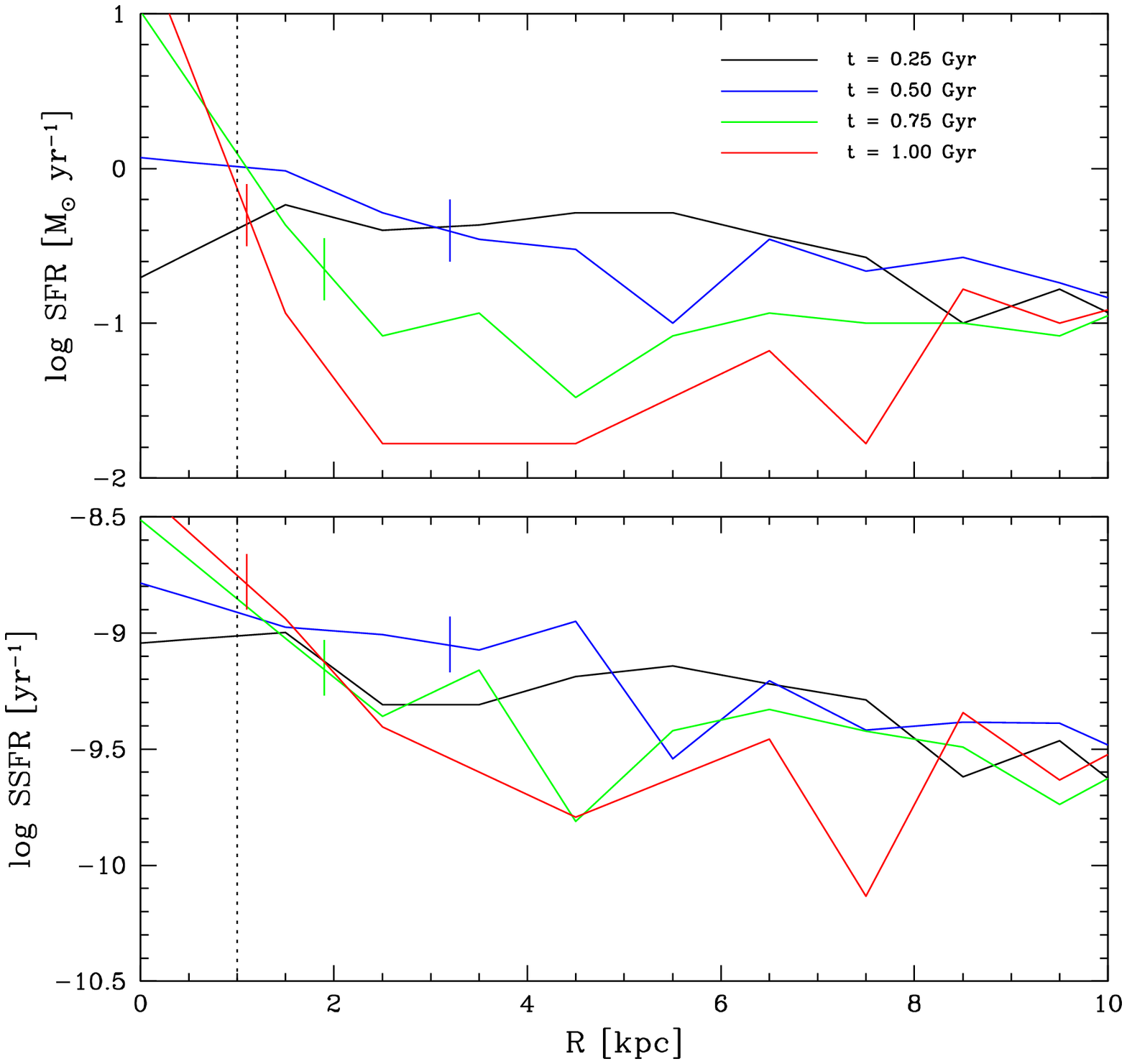}
\caption{Star formation rate (top panel) and specific star formation rate
(bottom panel) versus radius, for Run F. 
The different curves correspond to different epochs, as indicated.
Vertical color lines indicate the half-length of the
gaseous component of the bar.
The dotted black line indicates the radius of the central region.}
\label{sfr_profile}
\end{center}
\end{figure}

Four processes can modify the mass of a particular element in the gas phase
inside
the central region: (1) gas flowing into the central region, (2) gas flowing out
of the central region, (3) gas being converted into stars, and (4) gas
being released by stars in the form of stellar winds and SNe ejecta
(we will refer to these two processes together as
{\it stellar outflows}),.
We now investigate the role of these various processes.
The top panel of Figure~\ref{fiber_OH2} shows the rate of change
of hydrogen mass $\MH$ (black lines and left axes) and oxygen mass
$\MO$ (red lines and right axes) inside the central region.
The middle panel of Figure~\ref{fiber_OH2} shows the combined effect of
matter flowing across the boundary of the central region (solid lines)
and matter exchange between the stellar and gas phase (dotted lines).
The bottom panel of Figure~\ref{fiber_OH2} shows the combined effect
of all processes. 

Between $t=0.2\,\Gyr$ and $t=1.1\,\Gyr$, there are large
amounts of gas both {\it moving in\/} and {\it moving out\/} of the central
region (solid lines in top panel of Fig.~\ref{fiber_OH2}).
This can be understood by examining the velocity field of the
gas inside the bar (Fig.~\ref{xy_vel}). At $t=0.5\,\Gyr$, the gas inside the
bar moves along an elongated, elliptical orbit. This orbit intersects the
edge of the central region, hence the gas is flowing across the boundary,
both moving in and moving out. This elongated vortex contracts with time,
and after $t\sim1.1\,\Gyr$, it is entirely contained inside the central region. 
At this point, the flux of gas through the edge is small, and 
further enrichment is caused mostly by stars in the central region
(bottom dotted lines in top panel of Fig.~\ref{fiber_OH2}).
Because of the overall contraction of the gas along the bar,
there is a net inflow of gas into the central region 
(middle panel of Fig.~\ref{fiber_OH2}). The rate of net inflow is quite 
irregular because the gas tends to become clumpy, but overall most of the
inflow takes place between $0.2\,\Gyr$ and $1.1\,\Gyr$. As the gas
get progressively enriched by SNe and stellar winds, the flux of
oxygen increases relative to the flux of hydrogen, causing a net increase
of \OH\ inside the central region.

The net effect of star formation, SNe, and stellar winds is a decrease in the
amount of gas, since some of the matter remains trapped in stellar remnants
or in stars whose lifetime exceeds the duration of the simulation.
Because gas returned to the ISM by SNe and winds has been enriched, the net
effect is again an increase in \OH. There is a delay between gas flowing into
the central region and gas being removed by star formation, the latter becoming
important only after $\sim0.7\,\Gyr$. As a result, the total mass of
hydrogen and oxygen in the central 
region increases with time until $t\sim1.0\,\Gyr$,
and then decreases afterward (bottom panel of Fig.~\ref{fiber_OH2}).

In Figure~\ref{fiber_2p}, we combine the top right panel of Figure~\ref{fiber}
with the middle panel of Figure~\ref{fiber_OH2}, to show the effect of these
various processes on the central metallicity. Between $t=0.3\,\Gyr$ and
$t=0.7\,\Gyr$, \OH\ increases from 0.12 to 0.22, {\it even though the effect
of the central stars is negligible}. This early enrichment is entirely caused
by metal-rich gas flowing into the central region. This gas was not enriched
by central stars, and was therefore enriched in other regions, prior to
falling into the central region. After $t=0.7\,\Gyr$, central star formation
becomes important, but until $t=0.96\,\Gyr$, gas flow across the boundary 
remains the dominant process. During this period, \OH\ increases from 0.22
to 0.32. Star formation starts to dominate only after $t=0.96\,\Gyr$,
and gas flow across the boundary becomes negligible after $t=1.18\,\Gyr$.
Overall, half of the enrichment (from $\OH=0.12$ to $\OH=0.32$) takes place
before $t\sim1.0\,\Gyr$, when net inflow of metal-rich gas
was either the only process or the dominant process taking place.
In Figure~\ref{fiber_Fe_2p}, we plot the corresponding results for
iron. This results are qualitatively similar to the ones shown in
Figure~\ref{fiber_2p}. Gas inflow dominates until $t\sim1\,\Gyr$.
Afterward, the dominant process is net removal of gas by star formation
and stellar outflows.

These results imply that the central region
does not evolve in isolation.
About half
of the metals that end up in the central region were formed somewhere else. 
To understand this, we need to examine the global history of star formation
in the galaxy. We identified all star formation events in the simulation
(that is, a gas particle turning into a star particle between two output
steps).
Figure~\ref{sfr_map2} shows the epoch and the distance
from the center where each event took place. The red line indicates the
radius of the central region. The blue curve
shows the extent of the gaseous component of the bar, and was calculated 
separately by visual inspection of the right-hand-side panels of 
Figure~\ref{xy} and similar panels at different times. The blue curve
forms a near-perfect envelope that encloses most star formation events.
This indicates that star formation is not limited to the central region.
Instead, stars are forming everywhere along the bar. At $t=0.5\,\Gyr$,
the gaseous component of the bar is about $6\,\Gyr$ in length, that is, 
three times the diameter of the central region. Most of the stars forming 
inside the bar at that time form outside of the central region. As the
gaseous component of the bar contracts, a larger fraction of the stars form
inside the central region, and at $t\sim1.1\,\Gyr$, the gaseous
component of the bar falls entirely inside the central
region. This corresponds to the time when the next flux of metal-enriched
gas across the boundary of the central region becomes negligible,
as seen in Figures~\ref{fiber_2p} and~\ref{fiber_Fe_2p}.

We calculated the SFR profiles and specific SFR (SSFR) profiles at different
times. The results are shown in Figure~\ref{sfr_profile}. The curves at
$0.50\,\Gyr$ are rather flat, indicating that star formation
is taking place all along the bar, at roughly the same 
rate.\footnote{At $t=0.25\,\Gyr$, the bar has not formed yet.} 
Afterward, the gaseous component of
the bar contracts, and the SFR and SSFR become centrally concentrated.
However, the fluctuations in SSFR at late time remain fairly small. The SFR
becomes centrally concentrated at late time mostly because the gas available
for forming stars becomes itself centrally concentrated.

Notice that star formation is not limited to the bar. All the dots on the
right of the blue curve in Figure~\ref{sfr_map2} correspond to stars
forming in the disk, mostly inside the spiral arms. The apparent ``void''
between 1 and $5\,\kpc$, which is also visible in the top red curve
in Figure~\ref{sfr_profile}, is simply a geometric effect; if stars are forming
uniformly over the surface of the disk, we expect the SFR to increase
linearly with radius.

\subsection{Changing the Resolution}

\begin{figure*}
\begin{center}
\includegraphics[width=5in]{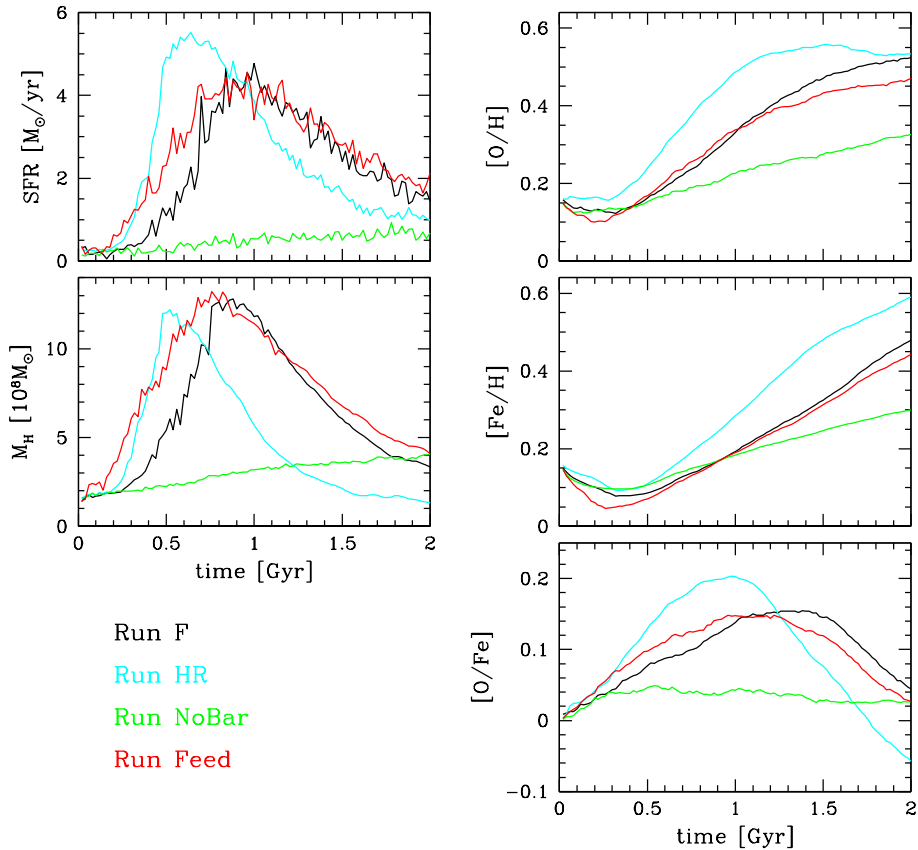}
\caption{Time-evolution of quantities inside the central region.
Top left panel: Star formation rate;
middle left panel: hydrogen mass;
right panels: \OH, \FeH, and \OFe\ in the gas phase, as indicated.
Black lines: Run F; cyan lines: Run HR; green lines: Run NoBar;
red lines: Run Feed.}
\label{allruns}
\end{center}
\end{figure*}

Figure~\ref{allruns} shows the evolution of the SFR, hydrogen mass, \OH, \FeH, 
and \OFe\ in the central region, for all runs. The blue and black lines show the
results for Runs HR and F, respectively, which only differ in the number
of particles used. As we see, increasing the resolution does not lead to
convergence. 
This is likely caused by the use of a fixed density threshold, $n_{\rm th}$, 
for star formation. As the resolution increases, the simulation is capable
of resolving finer structures, and more gas particles are eligible to reach
the threshold density more quickly. The finer structures and earlier 
starbursts could help developing more structures in the gas, which could
enhance the growth of the non-asymmetric structures of the stars, such as a 
bar and spiral arms. We should be able to adjust $n_{\rm th}$ and the other
parameters to match the lower-resolution run, but this is not our goal
in this paper. This comparison run demonstrates that the timescale of
the evolution of chemical and dynamical properties is sensitive to the
simulation set-up, but the evolution paths are qualitatively similar.
In future
work, we will continuously improve our simulations by calibrating these 
parameters against the various observational constraints \citep{rk12},
and explore also the initial conditions, such as the mass of
the disk and the gas mass fraction.

\begin{figure}
\begin{center}
\includegraphics[width=3.4in]{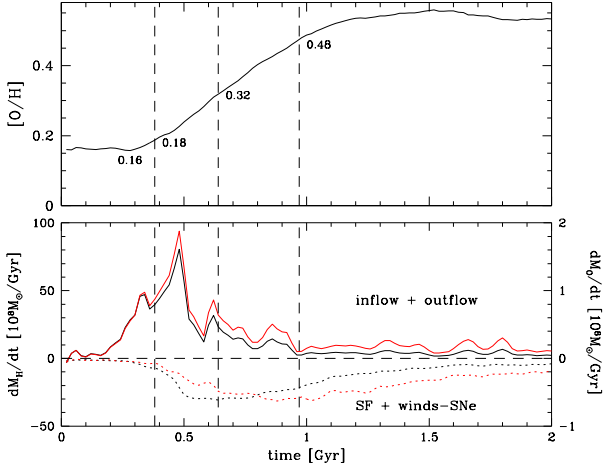}
\caption{Same as Figure~\ref{fiber_2p}, for Run HR.}
\label{fiber_2p_HR}
\end{center}
\end{figure}

Still, the results for the Runs F and HR are qualitatively similar. The
SFR and the hydrogen mass \MH\ initially increase, reach a peak, and then 
decrease. The SFR
peak is reached earlier in Run HR, at $t=0.7\,\Gyr$ instead of
$1.0\,\Gyr$. Looking at the right panels in Figure~\ref{allruns}, we see
interesting differences. In Run HR, \OH\ reaches a peak at $t\sim1.5\,\Gyr$, 
and starts to decrease. 
The plot of \OFe\ versus time shows that after $t=1\,\Gyr$,
Type~Ia SNe and mass loss from the intermediate-mass stars start 
to dominate. The stellar mass ejecta do not produce much oxygen, but
they produce a lot of hydrogen. Hence, the decrease in \OH\ is not
caused by oxygen depletion, but rather by dilution: \OH\ goes down because
the amount of hydrogen goes up relative to the amount of oxygen. 
In Run F, mass loss from intermediate-mass stars takes over much
later, and for that reason \OH\ keeps increasing until the end of the
simulation. Figure~\ref{fiber_2p_HR} shows the evolution of \OH, \MH,
and \MO\ in the central region, for Run HR. The vertical dashed lines indicate 
the same particular epochs as in Figure~\ref{fiber_2p}: star formation
becoming significant, equal contribution from inflow and stars,
and inflow becoming insignificant. 
Comparing with Figure~\ref{fiber_2p}, 
we see that the period during
which inflow + outflow of gas dominates is much shorter, extending
from $t=0.20\,\Gyr$ to $t=0.63\,\Gyr$. Also, star formation and stellar
outflows
become important earlier, around $t=0.38\,\Gyr$. But qualitatively,
the results are the same for Runs F and HR: an early period dominated by
gas flowing in and out of the central region, during which \OH\ increases 
from 0.16 to 0.32, followed by a late period dominated
by star formation and stellar outflows, 
during which \OH\ increases to a maximum of 0.56
before falling.

\subsection{Barred versus Unbarred Galaxy}

\begin{figure}
\begin{center}
\includegraphics[width=3.4in]{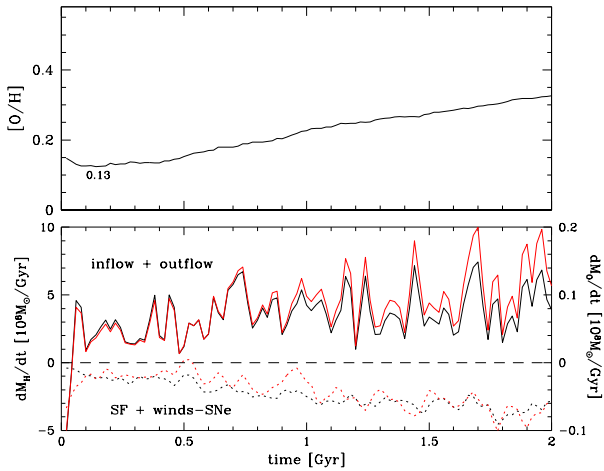}
\caption{Same as Figure~\ref{fiber_2p}, for Run NoBar.
Notice the difference in vertical scales between the bottom
panels of the two figures.}
\label{fiber_2p_NoBar}
\end{center}
\end{figure}

The green lines in Figure~\ref{allruns} show the results for Run NoBar.
In the absence of a bar, there is very little gas moving into the central
region, and both the SFR and \MH\ increase very slowly, resulting is smaller
changes in \OH, \FeH, and \OFe\ compared to the fiducial run.
Interestingly, at $t=2\,\Gyr$, the mass of hydrogen in the central
region is almost the same in Runs F and NoBar, though it gets to that value
in totally different ways. 
In Run F, the bar channels large
quantities of gas into the central region, but most of the gas is converted
into stars. In Run Nobar, the SFR is much lower. The SFR
increases as \MH\ increases,
but never gets high enough to result in a decrease in \MH.
Metal enrichment is reduced compared to Runs F and HR, and interestingly
\OFe\ is nearly constant.
Figure~\ref{fiber_2p_NoBar} shows the evolution of \OH, \MH,
and \MO\ in the central region, for Run NoBar.
The results are totally different than the ones for Run F (Fig.~\ref{fiber_2p}).
The rates are about an order of magnitude smaller, for all processes.
We do not find an early period dominated by inflow + outflow and a late
period dominated by star formation + stellar outflows. 
Instead, all processes are taking
place simultaneously, and all rates increase roughly linearly with time.
Inflow + outflow dominate at all times (bottom panel), indicating that the
increase in \OH\ and \FeH\ seen in the central region is mostly caused
by gas being enriched outside of the central region and moving in afterward.
In Runs F and HR, the gaseous component of the bar contracts and eventually
falls entirely inside the central region, shutting down inflow. This does not 
happens in the absence of a bar. Gas flows slowly and steadily into the 
central region all the way to the end of the simulation.

\subsection{Stronger Feedback}

\begin{figure}
\begin{center}
\includegraphics[width=3.4in]{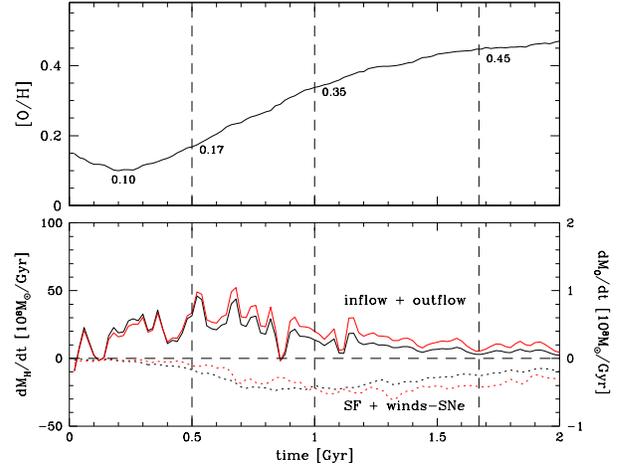}
\caption{Same as Figure~\ref{fiber_2p}, for Run Feed.}
\label{fiber_2p_Feed}
\end{center}
\end{figure}

The red lines in Figure~\ref{allruns} show the results for Run Feed.
The results are very similar to Run F. Increasing the stellar wind and
SNe feedback by an order of magnitude has virtually no effect on
the flow of the gas, indicating that this flow is still dominated
by gravitational dynamics. The only real difference is the central SFR and
central hydrogen mass, which start to rise earlier, but eventually reach
the same peak values at the same time. Having nearly the same SFR in
Runs F and Feed likely results from competing effects. A larger feedback
results in compression and reheating of the ISM, the former process favoring
star formation and the second process inhibiting it. Still, it is somehow
surprising that such a large increase in feedback makes so little difference. 
Figure~\ref{fiber_2p_Feed} shows the evolution of \OH, \MH,
and \MO\ in the central region, for Run Feed.
Again, the results are similar to those for Run F. Gas inflow dominates
up to $t=1\,\Gyr$, and during that period \OH\ increases from
0.10 to 0.35. Afterward, star formation and 
stellar outflows dominate, and \OH\ increases
up to 0.47.

\subsection{The Effect of Inclination and Aperture}

So far, we have defined the central region as a cylinder of diameter $2\,\kpc$
centered on the rotation axis of the galaxy. Observationally, this corresponds
to a disc galaxy seen face-on. In practice, face-on galaxies are fairly rare.
In this section, we redefine the central region as a cylinder of diameter 
$2\,\kpc$, not centered on the rotation axis, but rather on the line-of-sight.
We consider galaxies with a $45^\circ$ inclination. This is large enough to
potentially have an effect, while still small enough to allow for a visual
identification of the bar. 

\begin{figure*}
\begin{center}
\includegraphics[width=5.in]{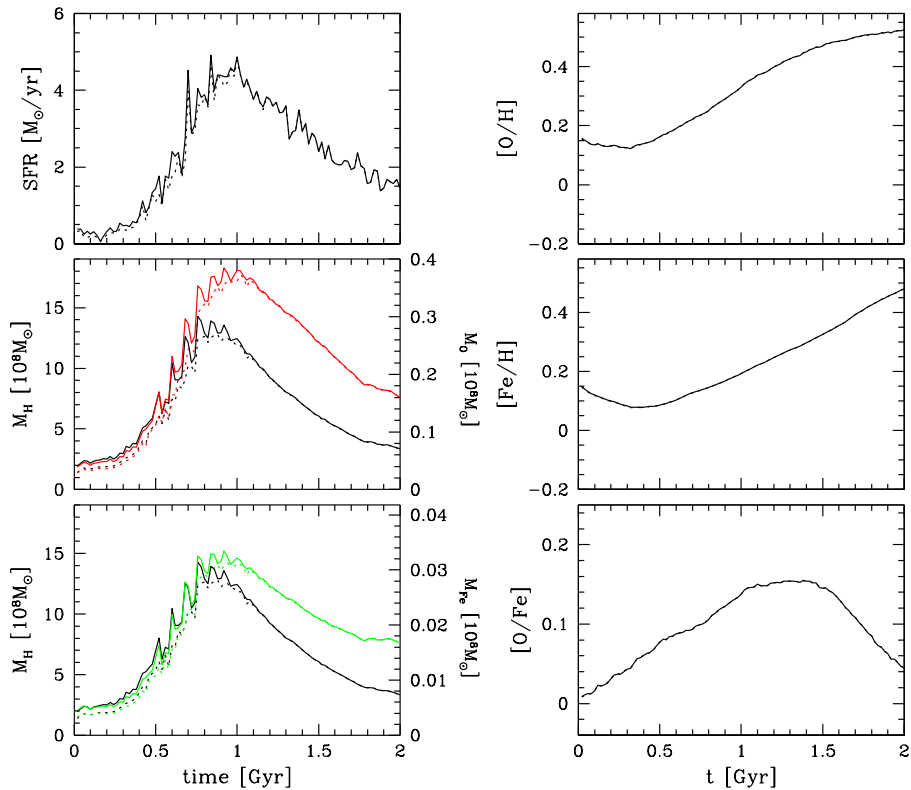}
\caption{Time-evolution of quantities inside the central region, for Run F.
Solid lines: $45^\circ$ inclination; dotted lines: no inclination.
Top left panel: Star formation rate;
middle and bottom left panels: hydrogen mass (black lines and left axes),
oxygen mass (red line and right axis in middle panel), and iron mass
(green line and right axis in bottom panel);
right panels: \OH, \FeH, and \OFe\ in the gas phase, 
as indicated. On these three panels,
the solid and dotted curves differ by only a few percents, and are
visually indistinguishable.}
\label{angle}
\end{center}
\end{figure*}

Figure~\ref{angle} shows the results for Run F. For comparison, we also
plot as dotted lines the results with no inclination, copied from
Figure~\ref{fiber}. They are very similar. In particular, \OH, \FeH,
and \OFe\ differ by less than 2\%, making the curves indistinguishable.
Most of the gas in the central region is concentrated in the inner 
$1\,\kpc$, so rotating a region of size $2\,\kpc$ only produces a small
edge effect. Therefore, our results are essentially insensitive to 
inclination.

\begin{figure}
\begin{center}
\includegraphics[width=3.4in]{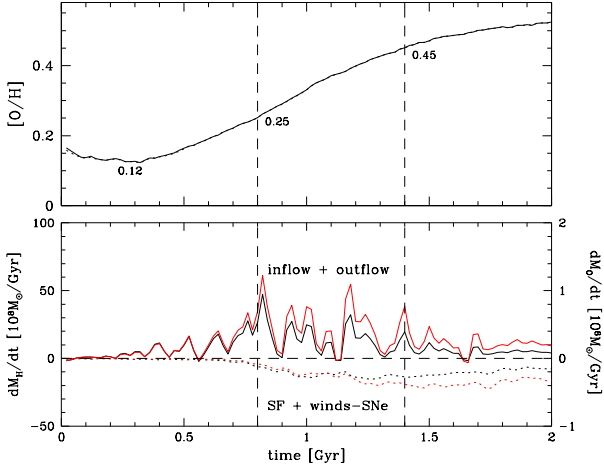}
\caption{Same as Figure~\ref{fiber_2p}, for Run F and
a central region of diameter $1\,\kpc$. Dotted line in top panel
shows the evolution of \OH\ for an aperture of $2\,\kpc$.}
\label{fiber_R05_2p}
\end{center}
\end{figure}

We have considered an aperture of $2\,\kpc$, because it corresponds
to the typical size of the fibre in the SDSS. But of course the actual
aperture is redshift-dependent. We now go back to Run F with no inclination
(face-on galaxy), and consider the effect of varying the aperture.
Figure~\ref{fiber_R05_2p} shows the evolution of \OH, \MH,
and \MO\ in the central region, for an aperture of diameter $1\,\kpc$.
Comparing with Figure~\ref{fiber_2p}, we find that reducing the aperture
does not qualitatively change the results. We still have an early phase
when gas inflow dominates, and a later phase when star formation and
stellar outflows 
dominate. However, everything is shifted to later times. The effect
of star formation and stellar outflows
become important at $t=0.8\,\Gyr$ (compared to $0.7\,\Gyr$ for the
$2\,\kpc$ aperture), the effects of gas inflow and stars are comparable
at $t=1.4\,\Gyr$ (compared to $0.96\,\Gyr$ for the $2\,\kpc$ aperture),
and at the end of the simulation, gas inflow is still important. The gaseous
component of the bar contracts with time, but it eventually reaches a minimum 
size which roughly corresponds to the aperture size of $1\,\kpc$
(see Fig.~\ref{sfr_map2}).

\begin{figure}
\begin{center}
\includegraphics[width=3.4in]{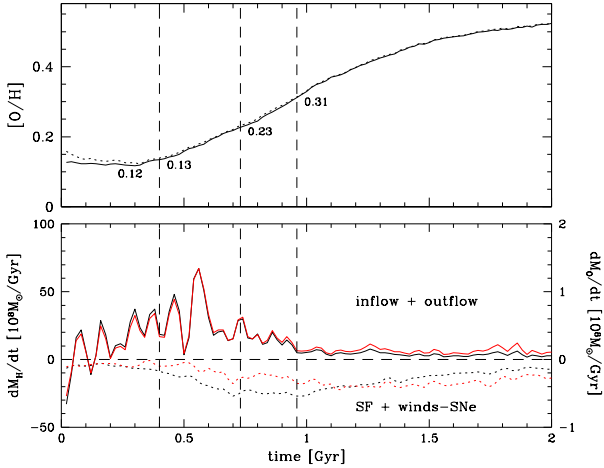}
\caption{Same as Figures~\ref{fiber_2p} and~\ref{fiber_R05_2p}, 
for Run F and a central region of diameter $5\,\kpc$. Dotted line in top panel
shows the evolution of \OH\ for an aperture of $2\,\kpc$.}
\label{fiber_R25_2p}
\end{center}
\end{figure}

Figure~\ref{fiber_R25_2p} shows the results for an aperture of diameter
$5\,\kpc$, which encloses most of the bar for the full duration of the
simulation. This time everything is shifted
to earlier times. The effects of star formation and stellar outflows
become important at $t=0.4\,\Gyr$, the effects of
gas inflow and stars are comparable at $t=0.73\,\Gyr$,
and gas inflow becomes negligible at $t=0.96\,\Gyr$, when the gaseous 
component of the bar falls entirely inside the central region.
Interestingly, the evolution of \OH\ is very insensitive
to the choice of aperture (top panels in Figs.~\ref{fiber_R05_2p}
and~\ref{fiber_R25_2p}), even though the timing of the physical processes
responsible for the evolution of \OH\ strongly depends on the choice
of aperture.

\section{DISCUSSION}

The classic picture of gas flowing along the bar,
producing a central starburst, and enriching the gas in the central region,
turns out to be
too simplistic. Several physical processes complicate this picture. 
The gas falling toward the central region tends to form an
elongated inner orbit
inside the stellar bar. Until this orbit contracts sufficiently to be
entirely contained inside the central region, this region does not evolve in
isolation. The dominant process during this period is not
star formation + stellar outflows in the central region, but rather exchange
of gas between the central region and the rest of the galaxy. The amount of
gas flowing in and out of the central region greatly exceeds the amount of
gas involved in star formation and stellar outflows, even when the central
SFR reaches its maximum value. {\it This implies, in particular, that 
metal-enrichment in the central region cannot be viewed as a 
mere consequence of star formation in the same region.} Stars formed in the 
central region can end up enriching the gas outside of the central region,
and vice-versa.

For instance, consider a star particle forming in the central region.
This star particle represents collectively an ensemble of stars
which have a range of initial masses, and therefore a range of lifetimes.
The production of metals by these stars and their deposition into the ISM will
therefore be extended in time\footnote{Incidently, this explains why
the abundance ratios vary much more
smoothly than the SFR.}. As the star particle
moves along the length of
the bar on an elongated orbit, it will deposit metals over the
entire length of the bar. The same is true for star particles forming in
the bar but outside of the central region: they will deposit metals over
the entire length of the bar, including in 
the central region. Furthermore, the
gas itself follows elliptical orbits, and this contributes to spreading metals
over the length of the bar.

\begin{figure}
\begin{center}
\includegraphics[width=3.4in]{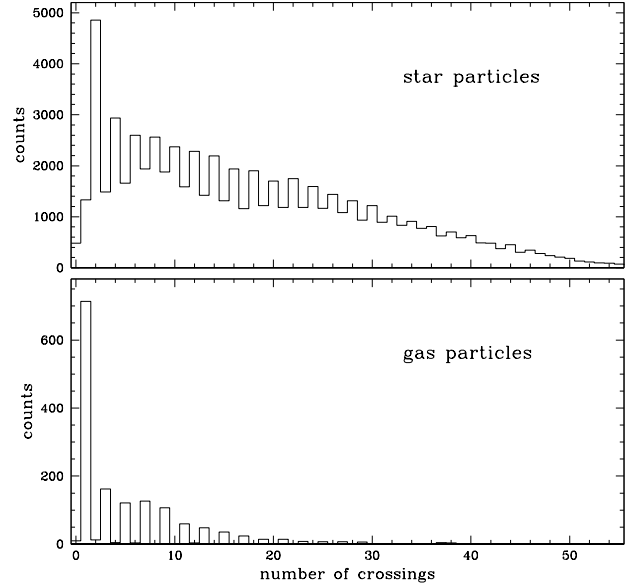}
\caption{Histograms of the number of times gas and star
particles have crossed
the boundary of the central region between $t=0\,\Gyr$ and $t=2\,\Gyr$,
for Run F. In the leftmost bin (zero crossings), we excluded
particles that were never inside the central region. The seesaw
pattern results from the fact that many particles start outside the central 
region at $t=0\,\Gyr$ and end up inside at $t=2\,\Gyr$, 
which implies an odd number of crossings.}
\label{inandout}
\end{center}
\end{figure}

To illustrate this, we calculated the number of times
each particle crossed the boundary of the central region, in Run F.
Figure~\ref{inandout} shows histograms
of the number of crossings, for star and gas particles. 
We excluded particles that were never inside the central region. Hence, the
leftmost bins (zero-crossings) represent particles that were located in
the central region in the initial conditions,
and remained in the central region throughout the simulation.

Star particles experience lots of crossings. The median value is 15 crossings.
When a particle moves along an elongated orbit that intersects the central
region, there are 4 crossings per orbit. Hence, 
about half of the star particles complete
between 1 and 4 full orbits. The histogram for gas particles is 
significantly different.
While the length of stellar component of the bar remains roughly constant
during the simulation, the gaseous component of the bar contracts with time.
As a result, most gas particles that cross the boundary of the central
region cross it only once. Most of the gas located in the bar will essentially
end up in the central region if we wait long enough. This implies that there
is no correlation between central star formation and central metal-enrichment,
but, in galaxies that are sufficiently evolved, there is a correlation between
{\it global\/} star formation and {\it central\/} metal-enrichment, in the 
sense that stars forming all over the length of the bar produce and release
metals which eventually end up in the central region.

\begin{figure*}
\begin{center}
\includegraphics[width=7in]{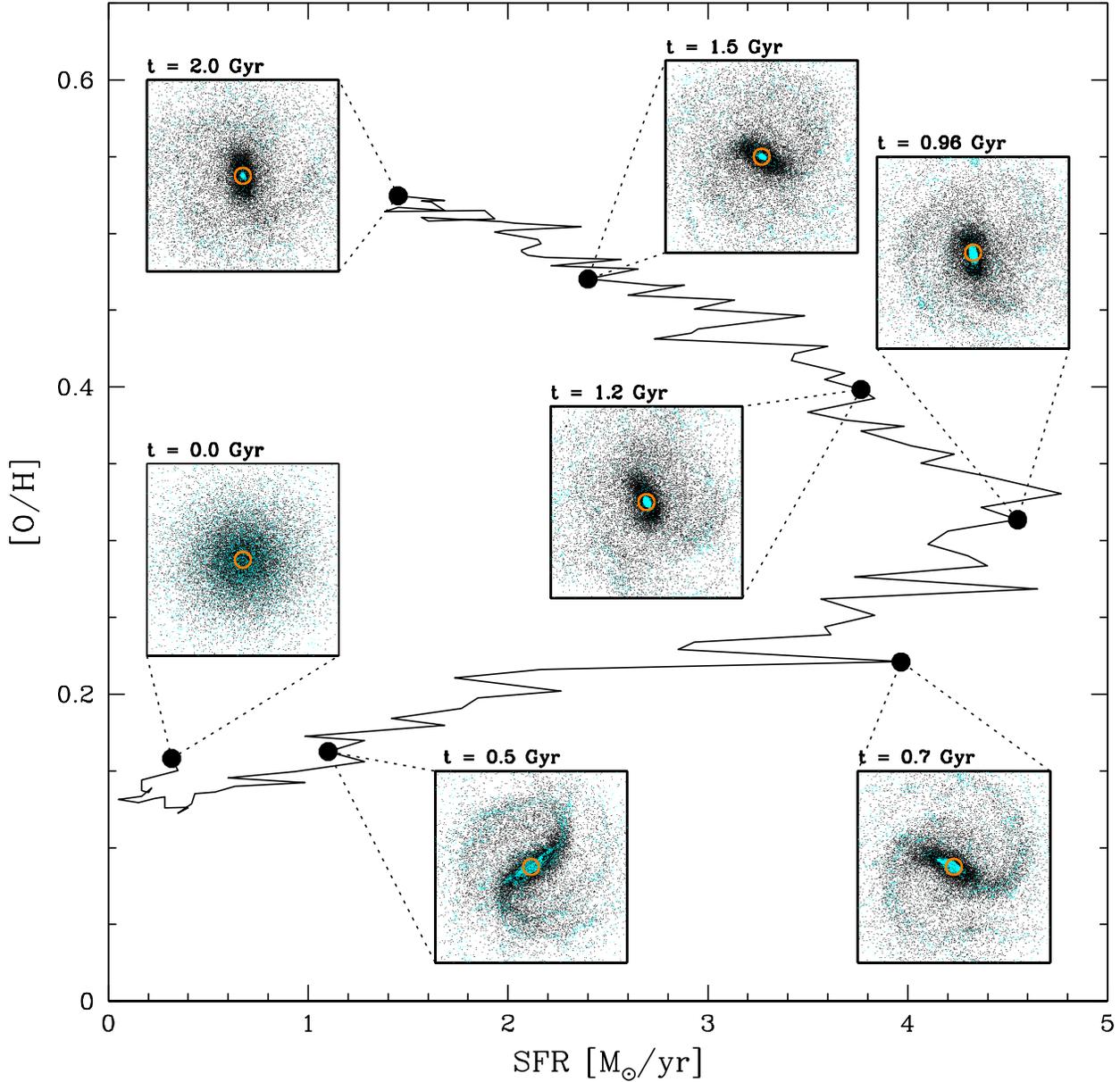}
\caption{\OH\ in the gas phase versus SFR, in the central region,
for Run F. Insets show configuration of star
particles (black dots), gas particles (cyan dots), and the boundary of the
central region (orange circles) at various times, as indicated. Each inset is 
$24\,\kpc\times24\,\kpc$. For clarity, only 1/6 of the particles are plotted.}
\label{sfr_OH}
\end{center}
\end{figure*}

We focused on the central $2\,\kpc$ because of its observational relevance.
It is interesting to see what an observer could infer about the
evolutionary stage of the galaxy, simply by observing the central region.
All quantities plotted in Figure~\ref{fiber} (SFR, \MH, \MO, \MFe,
\OH, \FeH, \OFe) vary non-monotonically with time. Hence, measuring
one of these quantities could not determine the current evolutionary
stage of the galaxy. Only a combination of measurements could achieve that.
To illustrate this, we plot the central value of \OH\ versus 
the central value of the SFR for Run F in
Figure~\ref{sfr_OH}. The line does not
intersect itself, indicating that each point on the line corresponds to
a specific epoch in the evolution of the galaxy. We have identified
seven particular epochs, and plotted the stellar and
gas distributions for each one.

\begin{itemize}

\item $t=0.0\,\Gyr$. Beginning of the simulation. The stellar and
gas distributions
are axisymmetric, the SFR is very low, and \OH\ has the initial value.

\item $t=0.5\,\Gyr$. This is the epoch when the bar is the strongest
and there is plenty of gas. 
The central SFR is still very low, and the dominant process responsible for
metal-enrichment of the central region is gas exchange between
the central region and the rest of the bar (Fig.~\ref{fiber_OH2}, top 
and middle panels; Fig.~\ref{fiber_2p}, bottom panel).

\item $t=0.7\,\Gyr$. The flux of gas in and out of the central region gets 
close to its maximum value, and star formation + stellar outflows start
to become important (Fig.~\ref{fiber_2p}, bottom panel).

\item $t=0.96\,\Gyr$. The SFR and gas mass in the central region reach their
maximum values (Fig.~\ref{fiber}, left panels).
The contribution of gas inflow and star formation + stellar outflows 
to metal-enrichment
of the central region are comparable (Fig.~\ref{fiber_2p}, bottom panel).
The orbit of the gas in the bar will soon be enclosed by the $2\,\kpc$
aperture (Fig.~\ref{xy_vel}, bottom left panel).

\item $t=1.2\,\Gyr$.
The orbit of the gas has fallen inside
the central region, and the flux of gas
across the boundary of the central region becomes insignificant.
(Fig.~\ref{fiber_2p}, bottom panel). 
From that point on, the central region evolves 
like a closed-box system, and star formation + stellar outflows are solely
responsible for further metal-enrichment of the central region.

\item $t=1.5\,\Gyr$. \OFe\ reaches its 
peak and starts decreasing (Fig.~\ref{fiber}, bottom right
panel), indicating a transition from
enrichment dominated by Type~II SNe to enrichment 
dominated by Type~Ia SNe.

\item $t=2.0\,\Gyr$. End of the simulation. The SFR steadily decreases as
the available gas in the central region
is depleted (Fig.~\ref{fiber}, left panels). 
\OH\ and \FeH\ both increase
at a constant rate (Fig.~\ref{fiber}, top right and middle right panels).

\end{itemize}

\section{SUMMARY AND CONCLUSION}

Our objective was to investigate the connection between the central SFR
and central metallicity in barred disc galaxies. In this initial study, we 
considered the particular case of a disc galaxy with a baryonic mass
of $5\times10^{10}\msun$ and initial stellar fraction of 80\%, in which a bar
forms by bar instability. We performed a series of four simulations.
We analyzed these simulations, focusing on the evolution of the SFR 
and metallicity in the central region of the galaxy, defined as the 
central kiloparsec (or $2\,\kpc$ in diameter), the approximate size of the
SDSS fibre.

In our first simulation, a strong bar of length $9\,\kpc$ composed of 
stars and gas forms around $t\sim0.5\,\Gyr$.
The stars and gas inside the bar do not lose all
their angular momentum. This prevents them from falling directly into the 
center. Instead, they form elongated orbits that go through the center of 
the galaxy. The orbits tend to contract with time, but the effect
is much more important for the gas than the stars. By the end of
the simulation, at $t=2\,\Gyr$, the stellar components of the bar
still has a length of order $7\,\kpc$, while the gaseous component
of the bar has contracted down to a length of less than $1\,\kpc$.

Before $t=0.5\,\Gyr$, when the bar has not yet formed, star formation is 
taking place mostly inside a radius of $5\,\kpc$,
where the gas is the
densest. At $t=0.5\,\Gyr$, when the bar forms, 69\% 
of the star formation takes place inside 
the bar, the remainder taking place inside the spiral arms. 
At this point, the gaseous component of the bar has a length
of about $7\,\kpc$, much larger than the size of the central region.
{\it Hence, most stars forming in the bar do not form in the central region.}
But as the gaseous component of the bar contracts, a larger fraction of that
gas is located in the central region, and the SFR in that region increases.
By $t=1.1\,\Gyr$, that gas is enclosed entirely inside the central region.
There is no more gas
flowing in, and the central SFR decreases as less gas is 
available to form new stars.

We tracked the evolution of the metal abundances in the gas phase,
in the central region.
We start the simulation with an initial metallicity 
gradient. This gradient is maintained in the
early stages of the simulation, when gas moves along circular orbits.
Once the bar forms, gas starts moving radially. 
The central metallicity initially decreases, and then
steadily increase until the end of the simulation.
Naively one could think that this increase in central metallicity is
caused by outflows from central stars. We decided to investigate this issue
in more detail,
by calculating
separately the contribution of the various process that
can change the
hydrogen mass, oxygen mass, and iron mass located in the central region
(Figs.~\ref{fiber_OH2}, \ref{fiber_2p}, and~\ref{fiber_Fe_2p}).
{\it The surprising result
is that the central metallicity increases significantly
before the contribution of central stars becomes significant.} 
Star formation and metal enrichment is taking place along the entire length 
of the bar. Because initially the central region is much smaller
than the bar, most of the metals are produced outside the central region.
As the gas moves along the bar on elliptical orbits, it moves in and 
out of the central region. But as these elliptical orbits contract with
time, there is a net flux of gas moving into the central region. This gas
was predominantly enriched by stars located outside the central region, and
early-on, this net influx of metal-enriched gas into the central region
dominates over the metals produced by the central stars.

Overall, about half of the increase in \OH\ and \FeH\ in
central region is caused by in-situ stellar enrichment.
The other half is caused by stars located outside the central region,
which produce metals that are
carried into the central region by large-scale gas flows along 
the bar.
The main conclusion is that
there is no direct connection between central SFR and central
metallicity, but there is a connection between global SFR and late-time,
central metallicity: stars forming over the entire length of the bar produce
metals that will eventually end up in the central region as the gaseous
component of the gas contracts.

Some recent IFU observations of barred galaxies appear to support
the conclusions of our simulations that star formation happens
along the length of the bar
\citep{cantinetal10,robertetal11,robertetal13}.
Histograms of the mass of the stellar populations
versus age for two barred galaxies, NGC4900 and NGC5430,
reveal that the stellar metallicity in the central region does not simply
increase gradually with time. Although the stellar metallicity
may be different than the gas 
metallicity, it suggests that the gas used to form the stars in
the central region may get enriched and diluted by an outside contribution
(like gas flowing along the bar).
Or the mixing process for the gas (between the time massive stars release
enriched elements and low mass stars release less enriched elements) involves
a complicated timing with the formation of new generation of stars.

We performed three more simulations. In one case, we increased the resolution
by a factor of 3, and in another case, we increased the amount of stellar 
wind and SNe feedback by a factor of 10. In both cases, we obtained results
that are quantitatively different, but qualitatively similar to our initial
run. In particular, we still identify an early phase when net gas inflow into
the central region is primarily responsible for increasing the central
metallicity, and a later phase when metal-enrichment by central stars
dominate. 
We also performed a simulation in which no bar forms. In that simulation,
there is very little gas flowing across the boundary of the central region,
and very few stars forming in the central region. But interestingly,
gas inflow is still the dominant process responsible of enriching the
gas in the central region.

In this paper, we focused on a particular set of initial conditions,
in order to investigate the relationships between the various physical
processes driving the evolution of the galaxy. In future work, we will
vary the initial conditions, specifically the mass of the galaxy and the
strength of the bar, with the goal of explaining the origin of the various
trends revealed by studies of barred galaxies in the SDSS.

\section*{acknowledgments}

We would like to thank Carmelle Robert for useful discussions.
Computations were performed on the Guillimin supercomputer at 
McGill University,
under the auspices of Calcul Qu\'ebec and Compute Canada. 
HM and SLE are supported by 
the Natural Sciences and Engineering
Research Council of Canada. HM is also supported by the Canada Research
Chair program.
DK acknowledges the support of the UK's
Science \& Technology Facilities Council (STFC Grant ST/H00260X/1).
HM is thankful to the Department
of Physics and Astronomy, University of Victoria, for its hospitality

\label{lastpage}

\end{document}